\documentclass[twocolumn,showpacs,prb,superscriptaddress]{revtex4}

\usepackage{graphicx}
\usepackage{color}
\usepackage{tabularx}
\usepackage{epsfig}
\usepackage{amsmath}
\usepackage{amssymb}
\usepackage{graphicx}
\usepackage{dcolumn}
\usepackage{bm}
\usepackage{wasysym}

\definecolor{grn}{rgb}{0,0,0.54}

\newcommand{\braket}[2]{\langle #1|#2\rangle}
\newcommand{\bra}[1]{\ensuremath{\langle #1 |}}
\newcommand{\ket}[1]{\ensuremath{| #1 \rangle}}
\newcommand{\expect}[1]{\langle \rangle}
\newcommand{\chif}{\chi_{\rm F}}
\newcommand{\chifg}{\chi_{\rm F}(g)}
\newcommand{\de}{{\partial^{2} E_{\rm 0}(g)}/{\partial g^{2}}}

\begin{document}


\title{Quantum Critical Scaling of Fidelity Susceptibility}

\author{A. Fabricio Albuquerque}
\author{Fabien Alet}
\author{Cl{\' e}ment Sire}
\author{Sylvain Capponi}
\affiliation{Laboratoire de Physique Th{\' e}orique, Universit{\' e} de Toulouse, UPS (IRSAMC), F-31062
Toulouse, France}
\affiliation{CNRS, LPT (IRSAMC), F-31062
Toulouse, France}


\date{\today}

\begin{abstract}

The behavior of the ground-state fidelity susceptibility in the vicinity of a quantum critical point
is investigated. We derive scaling relations describing its singular behavior in the
quantum critical regime. Unlike in previous studies, these relations are solely expressed in
terms of conventional critical exponents. We also describe in detail a quantum Monte Carlo scheme
that allows for the evaluation of the fidelity susceptibility for a large class of many-body systems and
apply it in the study of the quantum phase transition for the transverse-field Ising model on the square lattice.
Finite size analysis applied to the so obtained numerical results confirm the validity of our scaling
relations. Furthermore, we analyze the properties of a closely related quantity, the ground-state
energy's second derivative, that can be numerically evaluated in a particularly efficient way. The
usefulness of both quantities as alternative indicators of quantum criticality is examined.

\end{abstract}

\maketitle



\section{Introduction}
\label{sec:intro}

The quantity known as {\em fidelity} naturally appears in the field of quantum information
science as a way of determining the reliability of a given protocol for quantum information
transfer: the similarity between input $\ket{\Psi_{\rm in}}$ and output $\ket{\Psi_{\rm out}}$ states
can be quantified by simply computing the absolute value of the overlap between them,
$F = |\braket{\Psi_{\rm in}} {\Psi_{\rm out}}|$. Recently, after the pioneering work \cite{zanardi:06}
of Zanardi and Paunkovi{\' c}, and following the broader trend of cross-fertilization between
the fields of quantum information science and condensed matter physics,\cite{amico:08}
a number of studies have extended the scope of applicability of the concept of fidelity to
the study of quantum critical phenomena (for a review, see Ref.~\onlinecite{gu:08}).

The basic idea behind this so-called {\em fidelity approach} is simple. We consider
a general many-body Hamiltonian
\begin{equation}
    {\mathcal H}(g) = H_0 + g H_1 ~,
  \label{eq:hamiltonian}
\end{equation}
with ground-state $\ket{\Psi_{0} (g)}$, ${\mathcal H}(g)\ket{\Psi_{0}
(g)} = E_{\rm 0} (g) \ket{\Psi_{0} (g)}$. Since $\ket{\Psi_{0} (g)}$ undergoes major
changes in the vicinity of a quantum critical point (QCP) $g_{\rm c}$, we expect a sharp
drop in the fidelity,
\begin{equation}
F(g, d g) = | \braket{\Psi_{0} (g+ d g)}
{\Psi_{0} (g)} |~,
  \label{eq:fidelity}
\end{equation}
for small ($d g \rightarrow 0$) variations in $g$
close to $g_{\rm c}$. Therefore, by investigating the behavior of $F(g, d g)$
when couplings in the Hamiltonian are varied, one should be able to detect quantum criticality.
Besides its novelty, this approach is purely quantum
geometrical\cite{zanardi:07b} and therefore has the appeal that no {\em a priori} identification of order
parameters is required.

The concept of {\em fidelity susceptibility}\cite{you:07} $\chi_{\rm F}(g)$ naturally appears as the fidelity's
leading term in the limit $d g \rightarrow 0$,
\begin{equation*}
  F (g, d g\rightarrow 0) \simeq 1 - \frac{1}{2} \chi_{\rm F}(g) dg^{2}.
  \label{eq:chiF01}
\end{equation*}
(The linear term in $dg$ in the above expansion vanishes due to normalization of the wave-function --- alternatively it can be seen to arise from the fact that $F(g, d g)$ is maximum at $dg=0$ for any value of $g$.) The aforementioned drop in $F(g, d g)$ close to a QCP is thus associated
to a divergence in $\chif$ and the latter quantity may also be employed in
the study of quantum phase transitions. The situation here is reminiscent of the use of the specific heat
to detect {\em thermal} phase transitions: while the presence of singularities in the specific
heat for varying temperatures signals the location of finite-temperature critical points,
$\chi_{\rm F}(g)$ is a system's response to changes in the coupling constant $g$, whose divergencies
are associated to the occurrence of {\em quantum} phase transitions.

Although obviously some information is lost in going from $F(g, d g)$ to $\chifg$, and for instance it is currently not clear whether transitions of order higher than second can be detected by studying the latter, focusing on $\chifg$ has up to now proved to be a fruitful strategy. The main reason behind this is that it is possible to show\cite{you:07,venuti:07,chen:08b} that $\chi_{\rm F}(g)$ is closely related to more conventional physical quantities, such as imaginary-time dynamical responses. This is particularly advantageous since it allows one to rely on well established concepts and techniques from theoretical condensed-matter physics in order to draw conclusions on the properties of $\chifg$.  We follow this line of reasoning in this paper, in a twofold way.

First, we present the details of a recently introduced\cite{schwandt:09} quantum Monte Carlo (QMC) scheme that allows for the evaluation of $\chif$ for a large class of sign-problem-free models. This constitutes an important advance as the group of problems that can be studied within the fidelity approach is considerably enlarged, and additionally one benefits from the computational power of QMC methods. In particular, high-precision scaling analysis for models in dimensions higher than one is now possible: previous computations of $\chif$ for two-dimensional systems have relied on exact diagonalization (ED) techniques and were restricted to
small system sizes, something that precludes a precise determination of scaling dimensions in the vicinity of a QCP.

Second, by building upon the aforementioned relationship between $\chi_{\rm F}$ and response functions, we determine the scaling behavior of the fidelity susceptibility close to a QCP. The divergence of $\chi_{\rm F}(g)$ at $g_c$ is shown to be related to the critical exponent $\nu$ describing the divergence of the correlation length. In this way, and supported by the results obtained from the QMC simulations, we assess the validity of other scaling analysis for the divergence of $\chif$ that have recently appeared in the litterature.

Throughout the paper, we also analyze the properties of the ground-state energy's second derivative, $\de$.
This quantity is closely related to the fidelity susceptibility\cite{chen:08b} and, as explained in Sec.~\ref{sec:scaling},
scales in a way related to the scaling of $\chi_{\rm F}$ in the quantum critical regime.
Since the computation of $\de$ within QMC is much more efficient
than that of $\chif$, as discussed in Sec.~\ref{sec:SSE_d2E0}, it is important to address
the question of which of these quantities is best suited to the study of second-order
quantum phase transitions and of whether the current interest around the concept of fidelity susceptibility
is justified on practical grounds.

The paper is organized as follows. After reviewing basic concepts in the fidelity approach in Sec.~\ref{sec:def}, we analyze an extension of the concept of fidelity susceptibility to finite temperatures (a prerequisite for path-integral QMC simulations) and relate it to a more commonly employed metric for thermal states in Sec.~\ref{sec:finiteT}. We then perform a scaling analysis
of $\chif$ and $\de$ in Sec.~\ref{sec:scaling} and relate their scaling dimensions to conventional critical exponents. 

In Sec.~\ref{sec:sse}, we give a detailed account of the previously introduced\cite{schwandt:09} QMC scheme for calculating $\chif$ and further explore it (Sec.~\ref{sec:numerical}) in order to determine the scaling dimension of the fidelity susceptibility in one of the most paradigmatic models in the field of quantum phase transitions: the transverse-field Ising model (TIM) in two dimensions. Throughout the paper, concepts are illustrated by presenting results for the one-dimensional version of the TIM, which is exactly solvable.\cite{lieb:61,katsura:62,pfeuty:70}
A summary is given in Sec.~\ref{sec:conclusions} and important technical details are discussed in the Appendix.
Some of the results present in this paper have been first presented (by some of us) in Ref.~\onlinecite{schwandt:09}.

\section{Fidelity Susceptibility}
\label{sec:ChiF}

\subsection{Definition}
\label{sec:def}

We consider the limit of $d g \rightarrow 0$ and perturbatively calculate the overlap appearing in
Eq.~(\ref {eq:fidelity}) to leading order in $d g$. The fidelity susceptibility,
defined by Eq.~(\ref{eq:chiF01}), can easily be shown to read~\cite{you:07,zanardi:07b} 
\begin{equation}
    \chifg = \sum_{n \neq 0} \frac{|\bra{\Psi_{n} (g)} 
    H_1 \ket{\Psi_{0} (g)}|^{2}}{\left[ E_{n}(g) - E_{0}
    (g) \right]^{2}}~,
  \label{eq:ChiF_pert}
\end{equation}
in terms of the eigenbasis
\begin{equation}
    \sum_{n} \ket{\Psi_{n} (g)} \bra{\Psi_{n} (g)} = I
  \label{eq:spec_resolution}
\end{equation}
of ${\mathcal H}(g)$, ${\mathcal H}(g)\ket{\Psi_{n} (g)} = E_{n}(g) \ket{\Psi_{n} (g)}$.

Starting from Eq.~(\ref {eq:ChiF_pert}), one can relate\cite{you:07,venuti:07} $\chi_{\rm F}(g)$ to the
imaginary-time correlation function
\begin{equation}
    G_ {H_1}(\tau) = \theta(\tau) \left(\langle H_1(\tau) {H_1}(0) \rangle - \langle H_1 \rangle ^2\right) ~,
  \label{eq:G_tau}
\end{equation}
where $H_1(\tau) = e^{\tau {\mathcal H}} H_1 e^{-\tau {\mathcal H}}$, with $\tau$ denoting an imaginary time, $\theta(\tau)$ is the Heaviside step function and zero-temperature averages are defined
by $\langle {\cal O} \rangle =\bra{\Psi_{0} (g)} {\cal O} \ket{\Psi_{0} (g)}$ . Inserting Eq.~(\ref {eq:spec_resolution}) into this last equation and taking its Fourier transform we arrive to
\begin{equation}
\begin{split}
    \tilde{G}_ {H_1}(\omega) = \sum_{n \neq 0} \frac{|\bra{\Psi_{n} (g)} 
    H_1 \ket{\Psi_{0} (g)}|^{2}}{E_{n}(g) - E_{0}(g) +i\omega}~,
\end{split}
  \label{eq:G_omega}
\end{equation}
The similarity between Eqs.~(\ref {eq:ChiF_pert}) and (\ref {eq:G_omega}) is evident
and by simply performing a derivative, one can establish\cite{you:07,venuti:07}  the important result
\begin{equation}
    \chi_{\rm F}(g) = i\left. \frac{d\tilde{G}_ {H_1}(\omega)}{d \omega }\right|_{\omega=0}
    = \int_{0}^{\infty} d\tau \tau G_ {H_1}(\tau)~.
  \label{eq:ChiF_G_omega}
\end{equation}
This expression is remarkable for a number of reasons. First, it
relates $\chi_{\rm F}(g)$ to a dynamical response of the system to the
``driving term" $H_1$, evidencing its physical content. Second, as
discussed in detail in Sec.~\ref{sec:scaling}, Eq.~(\ref
{eq:ChiF_G_omega}) allows us to address the issue of the scaling
behavior of $\chi_{\rm F}(g)$. Finally, Eq.~(\ref {eq:ChiF_G_omega})
permits us to extend the definition of fidelity susceptibility to
finite temperatures (Sec.~\ref{sec:finiteT}) and therefore
constitutes an obvious starting point in devising a scheme to obtain
$\chi_{\rm F}(g)$ from path-integral QMC simulations.

Before proceeding, it is also instructive to consider the ground-state
energy's second-derivative, whose intimate relation to $\chi_{\rm
  F}(g)$ has been pointed out in Refs.~\onlinecite{chen:08b} and \onlinecite{venuti:08a}.  Motivated by
this close relationship, we define $\chi_{\rm E}(g) = -{\partial^{2}
  E_{\rm 0}(g)}/{\partial g^{2}}$ and, by using the eigenbasis of
${\mathcal H}(g)$, it is readily shown that
\begin{equation}
  \chi_{\rm E}(g) = -\frac{\partial^{2} E_{\rm 0}(g)}{\partial g^{2}} = 2 \sum_{n \neq 0}
   \frac{|\bra{\Psi_{n} (g)} {H_1} \ket{\Psi_{0} (g)}|^{2}}
   {E_{n}(g) - E_{0}(g)}~.
  \label{eq:d2_E0}
\end{equation}
Comparing Eqs.~(\ref {eq:G_omega}) and (\ref{eq:d2_E0}) it is straightforward to see that
\begin{equation}
    \chi_{\rm E}(g) = 2\tilde{G}_ {H_1}(\omega=0)
    = 2\int_{0}^{\infty} d\tau G_ {H_1}(\tau)~.
  \label{eq:d2_E0_omega}
\end{equation}
We can notice that the only important difference between Eqs.~(\ref{eq:ChiF_pert}) and
(\ref{eq:d2_E0}) is that in the former the denominator is squared: this is reflected by
the appearance of the $\tau$ factor in Eq.~(\ref{eq:ChiF_G_omega}), absent
in Eq.~(\ref{eq:d2_E0_omega}). One might thus expect $\chi_{\rm F}(g)$ to display a
more pronounced behavior around a QCP and therefore to be a better indicator of
quantum criticality, an observation put onto firmer grounds by the scaling analysis of
Sec.~\ref{sec:scaling}. Finally, similarly to the case of Eq.~(\ref{eq:ChiF_G_omega}),
the relation Eq.~(\ref{eq:d2_E0_omega}) can be used in order to extract $\chi_{\rm E}(g)$
from QMC simulations, as we discuss in Sec.~\ref{sec:SSE_d2E0}.  

\subsection{Finite Temperature}
\label{sec:finiteT}

\begin{figure}
  \begin{center}
    \includegraphics*[width=0.3\textwidth,angle=270]{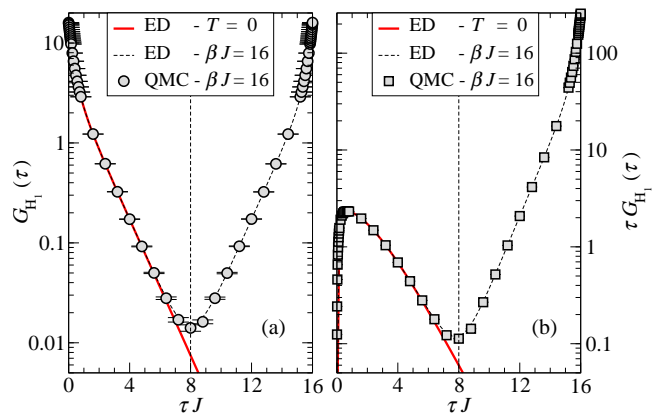}
   \end{center}
   \caption{(Color online) Correlation functions (a) $G_{H_1}(\tau)$ and
     (b) $\tau G_{H_1}(\tau)$, as a function of $\tau$, for finite
     inverse temperature $\beta J=16$ for the $d=1$ TIM on a
     chain with $L=16$ sites and $h/J=1$ (see
     Sec.~\ref{sec:transverse}). Data are generated by using the QMC method
     detailed in Sec.~\ref{sec:sse} and fixing the parity to the
     $P=+1$ sector (Sec.~\ref{sec:parity}). Exact Diagonalization (ED) data with fixed parity
     $P=+1$ are also shown both for finite and zero temperatures.}
  \label{fig:gtau_periodic}
\end{figure}

Before discussing how to extend the definition of fidelity susceptibility $\chi_{\rm F}(g)$ to finite temperatures
($T=1/\beta$) it is instructive to consider first the similar extension for $\chi_{\rm E}(g)$. From
Eq.~(\ref{eq:d2_E0_omega}), one obtains the finite-$T$ generalization
\begin{equation}
    \chi_{\rm E}(g,\beta) = 2\int_{0}^{\beta/2} G_ {H_1}(\tau)d\tau ~,
  \label{eq:ChiE_T01}
\end{equation}
where now $G_ {H_1}(\tau)$ is still defined by Eq.~(\ref{eq:G_tau})
but with {\em thermal} averages, $\langle {\cal O} \rangle = Z^{-1}
\mathrm{Tr } \left[ \exp(-\beta {\cal H}) {\cal O} \right]$ replacing
ground-state expectation values [$Z=\mathrm{Tr} \left\{ \exp(-\beta {\cal H})\right\}$
is the partition function]. An important subtlety is apparent here: notice that the
upper integration limit in the above expression is $\beta/2$, instead
of $\beta$. The underlying reason is that, within the path-integral
formalism used in QMC simulations, periodic boundary conditions are
implied along the imaginary time direction. Connected physical
correlation functions, such as $G_ {H_1}(\tau)$, are periodic along
the $\tau$-direction, with period $\beta$. This is illustrated in
Fig.~\ref{fig:gtau_periodic}(a) for the one-dimensional TIM (see
Sec.~\ref{sec:transverse}) on a chain with $L=16$ sites and $\beta
J=16$, for $h/J=1$: we see that $G_ {H_1}(\tau)$ is symmetric around
$\beta/2$ (vertical dashed line) and decays to zero at $\tau
\rightarrow \beta/2$ for large enough $\beta$, a trend already
noticeable in Fig.~\ref{fig:gtau_periodic}(a) where data for the
relatively high temperature $\beta J = 16$ are displayed. Therefore,
in the present case we have $\chi_{\rm E}(g) = 2 \int_{0}^{\beta/2} G_
{H_1}(\tau) d\tau = \int_{0}^{\beta} G_ {H_1}(\tau) d\tau$.

The definition of fidelity susceptibility [Eq.~(\ref{eq:ChiF_G_omega})] can be extended to finite temperatures in a similar way
\begin{equation}
    \chi_{\rm F}(g,\beta) = \int_{0}^{\beta/2} \tau  G_ {H_1}(\tau) d\tau~.
  \label{eq:ChiF_T01}
\end{equation}
An important difference appears though: the aforementioned properties
of $G_ {H_1}(\tau)$ are {\em not} shared by the function $\tau G_
{H_1}(\tau)$, since the pre-factor $\tau$ destroys the periodicity
along the imaginary-time direction This is illustrated in
Fig.~\ref{fig:gtau_periodic}(b), again using the one-dimensional TIM
as an example. In particular, $\int_{0}^{\beta/2} \tau G_ {H_1}(\tau)
d\tau \neq \int_{\beta/2}^{\beta} \tau G_ {H_1}(\tau) d\tau$ and,
therefore, in order to ensure that $\chi_{\rm F}(g,\beta)$ converges
correctly to its zero temperature limit, $\lim_{\beta\rightarrow
  \infty} \chi_{\rm F}(g,\beta)=\chifg$, one must cut the integral at
$\beta/2$. This has important implications for the QMC evaluation
that is now made possible by Eq.~(\ref{eq:ChiF_T01}), as clarified in
Sec.~\ref{sec:SSE_ChiF}.

While the just discussed generalization of $\chif$ to finite $\beta$ [Eq.~(\ref{eq:ChiF_T01})]
has been introduced for computational purposes, it is possible to relate
it to more commonly used metrics for thermal (mixed) quantum states,\cite{zanardi:07} as we
discuss in what follows.

\subsubsection{Bures Metric}
\label{sec:bures}

The so-called Uhlmann fidelity generalizes the concept of fidelity
[Eq.~(\ref{eq:fidelity})] to the case of mixed states. For density matrices $\rho_{\rm A}$ and
$\rho_{\rm B}$, it is defined as\cite{uhlmann:76}
\begin{equation*}
    {\mathcal F} (\rho_{\rm A},\rho_{\rm B}) = {\rm Tr} \sqrt{\rho_{\rm A}^{1/2} \rho_{\rm B} \rho_{\rm A}^{1/2}}~,
  \label{eq:uhlmann}
\end{equation*}
and has an associated metric $ds (\rho_{\rm A},\rho_{\rm B}) = \sqrt{2\left[ 1 - {\mathcal F} (\rho_{\rm A},\rho_{\rm B}) \right]}$, known as the Bures distance. We are interested in the case of {\em thermal} density matrices,
\begin{equation*}
\rho_{g} = \frac{1}{Z} \sum_{n} e^{-\beta E_{n}(g)}\ket{\Psi_{n}(g)}
                                                     \bra{\Psi_{n}(g)}~,
  \label{eq:thermal_rho}
\end{equation*}
expressed here in terms of the eigenbasis of ${\mathcal H}(g)$ [Eq.~(\ref{eq:spec_resolution})]. The concept of fidelity susceptibility [Eq.~(\ref{eq:ChiF_pert})] can then be extended to the finite-$T$ regime with the Bures metric $ds^{2}(g, \beta) = ds^{2} (\rho_{g}, \rho_{g + dg})$
for density matrices associated to infinitesimally close ($d g \rightarrow 0$) couplings in the Hamiltonian,
$g$ and $g + dg$. Following Zanardi {\it et al.},\cite{zanardi:07} one obtains the following expression for $ds^{2}(g, \beta)$
(for the sake of simplicity we omit the dependence on $g$ of the eigenvalues and eigenvectors in the remainder of this Section)
\begin{equation}
\begin{split}
    ds^{2}(g,\beta) & =  ds^{2}_{\rm cl}(g,\beta) \\  + &
     \sum_{n > m} \frac{\left| \bra{\Psi_{n}} {H_1} \ket{\Psi_{m}} \right|^{2}}
     {\left( E_{n} - E_{m}\right)^{2}}
    \frac{e^{-\beta E_{n}}}{Z}\frac{\left( 1 - e^{-2x} \right)^{2} } {1 + e^{-2x}} ~,
  \label{eq:Bures02}
  \end{split}
\end{equation}
where $x = \beta (E_m - E_n)/2$. One appealing feature in this expression is that it distinguishes ``classical'' and quantum
contributions. The ``classical'' term, $ds^{2}_{\rm cl}(g,\beta)$, is given by\cite{zanardi:07}
\begin{equation}
    ds^{2}_{\rm cl}(g,\beta) = \frac{\beta^{2}}{4} (\langle {H}^{2}_{1,\rm d} \rangle - 
    \langle H_{1,\rm d} \rangle^{2} )~.
  \label{eq:classical}
\end{equation}
Here, $H_{1,{\rm d}}$ denotes the diagonal elements of $H_1$ in
the eigenbasis Eq.~(\ref{eq:spec_resolution}), so that
\begin{equation*}
    \langle H_{1,{\rm d}} \rangle = \frac{1}{Z} \sum_{n} e^{-\beta E_{n}}
                  \bra{\Psi_{n}} H_1 \ket{\Psi_{n}}~.
  \label{eq:diagonal}
\end{equation*}
On the other hand, the second term
in Eq.~(\ref{eq:Bures02}) is of pure quantum origin and vanishes unless
$[\rho_{g}, \rho_{g + dg}] \neq 0$.\cite{zanardi:07}

\subsubsection{Relation between $\chi_{\rm F}(g,\beta)$ and $ds^{2}(g,\beta)$}
\label{sec:relation}

We now relate the two previously discussed finite temperature
extensions for the fidelity susceptibility, namely Eqs.~(\ref{eq:ChiF_T01}) and (\ref{eq:Bures02},
\ref{eq:classical}). We expand the trace for thermal averages and insert the
eigenbasis of ${\mathcal H}(g)$ [Eq.~(\ref{eq:spec_resolution})] in Eq.~(\ref{eq:ChiF_T01}) arriving to
\begin{widetext}
\begin{equation*}
    \chi_{\rm F}(g,\beta) =  \frac{1}{Z} \sum_{n,m} \int_0^{\beta/2} \! \! \! \! \! d\tau \tau 
    \left[ e^{-\beta E_n +\tau(E_n-E_m)} \left| \bra{\Psi_{n}} H_1 \ket{\Psi_{m}} \right|^{2} -
    \frac{e^{-\beta (E_n+E_m)}}{Z}   \bra{\Psi_{n}}H_1 \ket{\Psi_{n}}\bra{\Psi_{m}}H_1 \ket{\Psi_{m}}  \right]~.
\end{equation*}
\end{widetext}
The terms with $n=m$ in the first term in the integrand are independent of $\tau$ and can be regrouped with the
second term to yield $\frac{1}{2} ds^{2}_{\rm cl}(g,\beta)$. Performing the integration for the remaining terms, we finally
obtain
\begin{equation}
\begin{split}
    \chi_{\rm F}(g,\beta) & =  \frac{ds^{2}_{\rm cl}(g,\beta)}{2} \\  + &
     \sum_{n > m} \frac{\left| \bra{\Psi_{n}}H_1 \ket{\Psi_{m}} \right|^{2}}
     {\left( E_{n} - E_{m}\right)^{2}}
    \frac{e^{-\beta E_{n}}}{Z} {\left( 1 - e^{-x} \right)^{2} } ~,
  \label{eq:ChiF_T02}
  \end{split}
\end{equation}
where again we set $x = \beta (E_m - E_n)/2$. One can readily show
that in the limit $T \rightarrow 0$ both $\chi_{\rm F}(g,\beta)$
[Eq.~(\ref{eq:ChiF_T02})] and $ds^{2}(g,\beta)$
[Eq.~(\ref{eq:Bures02})] converge to the ground-state result of
Eq.~(\ref{eq:ChiF_pert}), as desired. This is illustrated for the TIM on
the square lattice in Fig.~\ref{fig:partition}, where data from QMC [for $\chi_{\rm F}(g,\beta)$]
and exact diagonalizations [for both $\chi_{\rm F}(g,\beta)$ and $ds^{2}(g,\beta)$] are displayed
(see discussion in Sec.~\ref{sec:pre_qmc}). On the other hand, the
high-temperature limit ($\beta \rightarrow 0$) yields $ds^{2}(g,\beta
\rightarrow 0) = 2\chi_{\rm F}(g,\beta \rightarrow 0) = \frac{\beta^2}{4}(\langle H_1^2\rangle -\langle H_1 \rangle^2)$.

In order to analyze the general case, we evaluate the ratio between $f(x) = (1-e^{-2x})^{2}/(1+e^{-2x})$,
that appears in Eq.~(\ref{eq:Bures02}), and $g(x)=(1-e^{-x})^{2}$ [from Eq.~(\ref{eq:ChiF_T02})].
We have $f(x)/g(x)=1 + 1/\cosh(x)$ and therefore $f(x)/2\leq g(x) \leq f(x)$.
Noting that  $ds^{2}_{\rm cl}(g,\beta)\geq 0$, we can conclude that
\begin{equation}
    \frac{1}{2} ds^{2}(g,\beta) \leq \chi_{\rm F}(g,\beta) \leq ds^{2}(g,\beta)~.
  \label{eq:bounds}
\end{equation}
These inequalities show that if $ds^{2}(g,\beta)$ diverges, then $\chi_{\rm F}(g,\beta)$
must also diverge and we conclude that both quantities are equally well suited for detecting criticality.
While in this paper we are interested in quantum phase transitions and thus focus on the
limit $\beta \rightarrow \infty$, it would also be of interest to investigate the finite-$T$ behavior
of $\chi_{\rm F}(g,\beta)$: as discussed in Ref.~\onlinecite{zanardi:07}, this quantity might be
able to detect thermal phase transitions and/or finite-$T$ signatures of quantum criticality.

\subsection{Scaling Behavior}
\label{sec:scaling}

We focus now on the issue of how the fidelity susceptibility $\chif$ behaves in the vicinity of a QCP
and start by reviewing the main results from the scaling analysis devised by Campos Venuti and Zanardi.\cite{venuti:07}

Starting from Eq.~(\ref{eq:ChiF_G_omega}) and following standard scaling arguments
(see for instance Ref.~\onlinecite{continentino:01}), they apply the scale transformation
$x' = sx$, $\tau' = s^z\tau$ ($z$ is the dynamic critical exponent), and arrive to the following relation
for the scaling of the fidelity susceptibility density:
\begin{equation}
    L^{-d}\chi_{\rm F} \sim |g - g_{\rm c}|^{\nu (d + 2z - 2\Delta_{H_1})}~.
  \label{eq:dim_chiF_02}
\end{equation}
Here, $L^{d}=N$ is the number of sites for a $d$-dimensional system. In what follows, we assume
that a second-order quantum phase transition takes place at a value $g_{\rm c}$ of the ``driving parameter"
$g$ and that the correlation length diverges in its neighborhood as $\xi \sim |g - g_{\rm c}|^{-\nu}$;
$\Delta_ {H_1}$ is the scaling dimension of the driving term $H_1 $ in Eq.~(\ref{eq:hamiltonian}):
${H_1}' = s^{-\Delta_ {H_1}} {H_1}$. Standard finite-size scaling arguments thus imply that
\begin{equation}
   {\chi_{\rm F}} \sim L^{-2(z - \Delta_{H_1})}
   \label{eq:venuti_scaling}
\end{equation}
for finite systems at criticality.

An alternative scaling analysis has been recently put forward by some of us in
Ref.~\onlinecite{schwandt:09}, where the simpler result
\begin{equation}
   {\chi_{\rm F}} \sim L^{2/ \nu}~,
   \label{eq:chif_scaling}
\end{equation}
has been derived. The appeal of the above scaling relation, as compared to Eq.~(\ref{eq:venuti_scaling}),
stems from the fact that it is not expressed in terms of the exponent $\Delta_ {H_1}$,
rarely dealt with in more conventional approaches to quantum critical phenomena. In fact, $\Delta_ {H_1}$ can
easily be related to the exponent $\nu$ (see Ref~\onlinecite{schwandt:09}). In what follows,
we provide an intuitive equivalent derivation of the result Eq.~(\ref{eq:chif_scaling}).

We start by remarking that at $T=0$ the free-energy density $f_{\rm H}$ reduces to the ground-state
energy density, ${L^{-d}}E_{\rm 0}$. Since, by definition, $f_{\rm H} \sim |g - g_{\rm c}|^{2 - \alpha}$
(see for instance Ref.~\onlinecite{continentino:01}), one readily obtains
\begin{equation}
  L^{-d}\chi_{\rm E} \sim |g - g_{\rm c}|^{-\alpha}~.
  \label{eq:dim_d2E0_02}
\end{equation}
This relation is not at all surprising since, as we can see from Eq.~(\ref{eq:d2_E0_omega}),
$\chi_{\rm E}(g) = - \partial^{2} E_{\rm 0}(g)/\partial g^{2}$ is similar to a ``zero-temperature specific heat''. 
Comparing the expressions for $\chi_{\rm E}$ [Eq.~(\ref{eq:d2_E0_omega})] and $\chi_{\rm F}$
[Eq.~(\ref{eq:ChiF_G_omega})], one sees that the only difference is the presence of an imaginary time
(or inverse energy) scale $\tau$ in the latter, that we naturally expect to scale as $|g - g_{\rm c}|^{-z \nu}$
in the critical regime.\cite{continentino:01} We thus arrive at 
\begin{equation}
    L^{-d}\chi_{\rm F} \sim |g - g_{\rm c}|^{-(\alpha + z \nu)} = |g - g_{\rm c}|^{-(2 - d \nu)}~,
  \label{eq:dim_chiF_03}
\end{equation}
where we have made use of the hyper-scaling formula $2-\alpha = \nu(d+z)$.
Therefore, we arrive to the conclusion that $L^{-d}\chi_{\rm F}$ diverges only when
$\nu < 2/d$.

On the other hand, by similarly inserting the hyper-scaling formula into
Eq.~(\ref{eq:dim_d2E0_02}), we obtain $L^{-d}\chi_{\rm E} \sim |g - g_{\rm c}|^{-2 + \nu(d+z)}$ and,
as anticipated in Sec.~\ref{sec:def}, conclude that $\chi_{\rm E}$ has a weaker
divergence than $\chi_{\rm F}$ at a critical point. Furthermore,
we find that the more stringent condition
$\nu < 2/(d + z)$ must be satisfied for $L^{-d}\chi_{\rm E}$ to diverge and
that there might be situations where only $L^{-d}\chi_{\rm F}$ displays a divergence at a critical point,
being in general a better indicator of quantum
criticality.

Next, by performing a finite-size scaling analysis, we conclude that
for the fidelity susceptibility per site we have
\begin{equation}
    L^{-d}\chi_{\rm F} \sim L^{\frac{2}{\nu} - d}~,
  \label{eq:dim_chiF_04}
\end{equation}
in agreement with the result Eq.~(\ref{eq:chif_scaling}) (Ref.~\onlinecite{schwandt:09}). 
For $L^{-d}\chi_{\rm E}$, we similarly obtain
\begin{equation}
  L^{-d}\chi_{\rm E} \sim L^{\frac{2}{\nu} - (d+z)}~.
  \label{eq:dim_d2E0_03}
\end{equation}
The validity of the scaling relations Eqs.~(\ref{eq:dim_chiF_04}) and (\ref {eq:dim_d2E0_03})
is confirmed by the analysis of the QMC data performed in Sec.~\ref{sec:results}. Finally, we note
that the scaling relations presented here have been independently found in the context of quantum
quenches,\cite{grandi:10,barankov:09,grandi:09} as discussed in Sec.~\ref{sec:conclusions}.

\section{Stochastic Series Expansion}
\label{sec:sse}

Despite the sign problem that precludes efficient simulations of
most fermionic/frustrated models, Quantum Monte Carlo methods are among the
most efficient tools to simulate quantum many-body
problems. Particularly useful for our purpose here is the QMC
formulation known as {\em Stochastic Series Expansion} (SSE),
developed by Sandvik and coworkers.\cite{sandvik:91,sandvik:99,
  syljuasen:02} Basically (for a detailed account the reader is
referred to Ref.~\onlinecite{syljuasen:02}), SSE relies on a power
series expansion of the system's partition function\cite{sse}
\begin{equation}
    Z = {\rm Tr} \left( e^{-\beta {\mathcal H}} \right) =\sum_{n=0}^{\infty} \sum_{\alpha} \sum_{S_{n}}
           \frac{\beta^{n}}{n!} \left\langle  \alpha \left|  \prod_{i=1}^{n} {\mathcal H} (b^{i})
           \right| \alpha \right\rangle.
  \label{eq:Z}
\end{equation}
Here $\left\{ \ket{\alpha} \right\}$ is any suitable basis and the system's Hamiltonian is typically a sum over local
operators: ${\mathcal H} = \sum_{b} {\mathcal H} (b)$, with $b$ labeling different local terms. For instance, $b$ may denote operators acting on different bonds of the lattice and/or diagonal versus non-diagonal operators. For our current purposes, it is convenient to choose a decomposition that respects the bipartition of Eq.~(\ref{eq:hamiltonian}), such that all terms appearing in $H_0$ are labelled by $b_0$ and those appearing in $g H_1$ by $b_1$ and we have $b \in \{b_0,b_1\}$. SSE configurations $(\alpha, S_{n})$, with operator strings
\begin{equation}
    S_{n} = \prod_{i=1}^{n} {\mathcal H} (b^{i})~,
  \label{eq:Sn}
\end{equation}
are then sampled, according to the statistical weight
\begin{equation*}
  W(\alpha, S_{n}) = \frac{\beta^{n}}{n!} \left\langle  \alpha \left|  \prod_{i=1}^{n} {\mathcal H}
                                    (b^{i}) \right| \alpha \right\rangle~.
  \label{eq:W}
\end{equation*}
Efficient update schemes such as the {\em directed loop} algorithm\cite{syljuasen:02,alet:05a} render the SSE technique
one of the most efficient QMC methods for quantum lattice models.

The general procedure for obtaining thermal averages within the SSE framework is discussed in
detail by Sandvik in Ref.~\onlinecite{sandvik:92}. The basic idea, supposing we are interested in
an observable ${\mathcal O}$, is to determine an {\em estimator} $O(\alpha, S_{n})$ such that
\begin{equation*}
    \left\langle {\mathcal O} \right\rangle_{W} = \frac{1}{Z} \sum_{n} \sum_{(\alpha, S_{n})} O(\alpha, S_{n})
    W(\alpha, S_{n})~.
  \label{eq:aver_O}
\end{equation*}
In what follows, we show how estimators for the fidelity susceptibility $\chi_{\rm F}(g,\beta)$
[Eq.~(\ref{eq:ChiF_T01})] and $\chi_{\rm E}(g,\beta)$ [Eq.~(\ref{eq:ChiE_T01})]
can be obtained from SSE QMC simulations.


\subsection{Fidelity Susceptibility}
\label{sec:SSE_ChiF}

First, we need to evaluate imaginary-time operator products of the form $\langle {H_1}(\tau)
{H_1}(0) \rangle$ appearing in the integrand of Eq.~(\ref{eq:ChiF_T01}) [cf.~Eq.~(\ref{eq:G_tau})].
These operators being part of the Hamiltonian, one trick consists in re-interpretating two of the
elements with label $b_1$ of the string Eq.~(\ref{eq:Sn}) as the operators to be measured. Following
Ref.~\onlinecite{sandvik:92}, we arrive to
\begin{equation}
\begin{split}
    & g^2 \left\langle {H_1}(\tau) {H_1}(0) \right\rangle =\\
    &   \sum_{m=0}^{n-2} \frac{(n-1)!}{(n-m-2)!m!}  \beta^{-n} (\beta-\tau)^{n-m-2}
    \tau^{m} \left\langle N_{gH_1}(m) \right\rangle_{W}.
  \label{eq:AtauA0}
\end {split}
\end{equation}
Here, $n$ is the length of the operator string $S_n$ [Eq.~(\ref{eq:Sn})] and
$N_{gH_1}(m)$ the number of times any two operators comprising $ g H_1$ appear in the strings
$S_{n}$ separated by $m$ positions. We discuss below how $N_{gH_1}(m)$ can be measured.  

The second term in Eq.~(\ref{eq:G_tau}) is obtained by a simpler
procedure\cite{sandvik:92} and is given by
\begin{equation}
    \left\langle {H_1} \right\rangle^{2} = \frac{1}{g^2\beta^{2}}
                           \left\langle N_{gH_1} \right\rangle^{2}_{W}~,
  \label{eq:Asq}
\end{equation}
where $N_{gH_1}$ is the total number of $g H_1$ operators in $S_{n}$.

Inserting the results Eqs.~(\ref{eq:AtauA0}, \ref{eq:Asq}) into Eq.~(\ref{eq:ChiF_T01}) and integrating
from $\tau=0$ to $\beta/2$ (taking into account the important multiplicative factor of $\tau$ in the
integrand), we finally arrive to the result
\begin{equation}
    \chi_{\rm F}(g,\beta) = \frac{1}{g^2} \sum_{m=0}^{n-2} \left[ A(m,n) \left\langle N_{gH_1}(m)\right\rangle_{W} \right]
                           -\frac{\left\langle N_{gH_1} \right\rangle_{W}^{2}}{8 {g^2}} ~,
  \label{eq:ChiF_QMC}
\end{equation}
with the coefficient
\begin{equation}
    A(m,n) = \frac{(n-1)!}{(n-m-2)!m!} \int_{0}^{1/2} \! \! \! \! \! d\tau \tau^{m+1} (1 - \tau)^{n-m-2}.
  \label{eq:integral}
\end{equation}
We show in Appendix~\ref{appendix} how this coefficient can be approximated very accurately by an analytical expression in the limit of $n\gg 1$.  

$N_{gH_1}(m)$  is conveniently extracted from the simulations in two steps. Firstly,  the string Eq.~(\ref{eq:Sn}) is traversed (for instance when performing diagonal updates; see Ref.~\onlinecite{sandvik:91}) and the positions $i$ where a local Hamiltonian ${\mathcal H}(b^i)$ appears with a label $b^i=b_1$ are recorded (there are in total $N_{gH_1}$ such operators).  Secondly, the histogram $N_{gH_1}(m)$ is generated by computing all distances $m$ between all previously recorded positions $i$. This step is the most demanding as it requires $N_{gH_1}(N_{gH_1}-1)/2$ operations. Note finally that the prefactor $1/g^2$ arises from the definition of  the fidelity susceptibility Eq.~(\ref{eq:ChiF_pert}) which does not include the coupling constant $g$, whereas the SSE decomposition used in Eq.~(\ref{eq:Z})  typically does.


\subsection{Ground-State Energy's Second Derivative}
\label{sec:SSE_d2E0}

The results Eqs.~(\ref{eq:AtauA0}, \ref{eq:Asq}) can also be used in order to directly evaluate
the ground-state energy's second derivative, relying on Eq.~(\ref{eq:ChiE_T01}) and extrapolating to the limit $\beta \rightarrow \infty$. 
The absence of the factor $\tau$ in Eq.~(\ref{eq:d2_E0_omega}) considerably simplifies the situation since the integration over $\tau$ can now always be performed exactly. In this way, we arrive to the simple result
\begin{equation}
     \chi_{\rm E} (g,\beta) = \frac{1}{g^2 \beta} \left[ 
     {\left\langle N_{gH_1}^{2} \right\rangle_{W} } - {\left\langle N_{gH_1} \right\rangle_{W} } 
     -{\left\langle N_{gH_1} \right\rangle_{W}^{2} } \right]~.
  \label{eq:d2E0_QMC}
\end{equation}
We stress that the computational cost for evaluating $\chi_{\rm E} (g,\beta)$
is much lower than the one required to obtain $\chi_{\rm F}(g,\beta)$: the estimator for the former
quantity in Eq.~(\ref{eq:d2E0_QMC}) simply requires counting the number of times the operators contained in
the ``driving term" $g H_1$ occur in the operator strings $S_{n}$. This is to be contrasted with the
computationally heavy task, specially in the limit of large lattice sizes and low temperatures, of computing
the histogram $N_{gH_1}(m)$ necessary in evaluating $\chi_{\rm F}(g,\beta)$
[cf.~Eqs.~(\ref{eq:ChiF_QMC}, \ref{eq:integral})].


\section{Numerical Simulations}
\label{sec:numerical}

\subsection{Transverse-Field Ising Model}
\label{sec:transverse}

\subsubsection{Definition}

The transverse-field Ising model (TIM) is perhaps the simplest model to display a
QCP and many key concepts in the theory of quantum critical phenomena have been
developed by analyzing its properties.\cite{sachdev:99a} 

The TIM Hamiltonian reads
\begin{equation}
    {\mathcal H}(h) = 	JH_{J} + hH_{h} = 
    				-J \sum_{\langle i,j \rangle} \sigma_{i}^{x} \sigma_{j}^{x}
                                     -h \sum_{i} \sigma_{i}^{z}~,
  \label{eq:tfim01}
\end{equation}
where ${\langle i,j \rangle}$ denotes nearest-neighbor sites on a $d$-dimensional lattice
and $\sigma_{i}^{x, z}$ are Pauli matrices attached to the site $i$. We set the energy scale
by henceforth fixing $J=1$.

At zero temperature and in any dimension, a quantum phase transition occuring at a field $h_{\rm c}$
separates a ferromagnetic phase for low fields $h<h_{\rm c}$ from a polarized one for $h>h_{\rm c}$,
with spins aligning along the field direction. The quantum phase transition in $d$ dimensions
belongs to the universality class of the finite-temperature phase transition of the classical Ising
model in $d+1$ dimensions and has a dynamic critical exponent $z=1$.\cite{sachdev:99a}

The TIM is exactly solvable in the one-dimensional case.\cite{lieb:61,katsura:62,pfeuty:70} The QCP is located at
$h_{\rm c}=1$ and most observables, including $\chi_{\rm F}$,\cite{gu:08} can be computed analytically.
We make use of these exact results in establishing the validity of the QMC method discussed in
Sec.~\ref{sec:sse}, for instance in Sec.~\ref{sec:parity}, where the issue of
parity is discussed. On the other hand, the TIM is not solvable in two dimensions
and has been investigated mainly through means of numerical techniques.\cite{he:90,oitmaa:91,croo:98,
rieger:99,hamer:00,bloete:02} The most precise estimate for the location of the QCP, $h_{\rm c}=3.04438(2)$,
has been obtained from a QMC approach.\cite{bloete:02}

Before proceeding, we remark that a scaling analysis for 
the fidelity susceptibility and the second derivative of the ground-state energy for the TIM on the
square lattice has been recently performed by Yu and collaborators in Ref.~\onlinecite{yu:09}.
Results are compared in Sec.~\ref{sec:conclusions}.


\begin{figure}
  \begin{center}
    \includegraphics*[width=0.3\textwidth,angle=270]{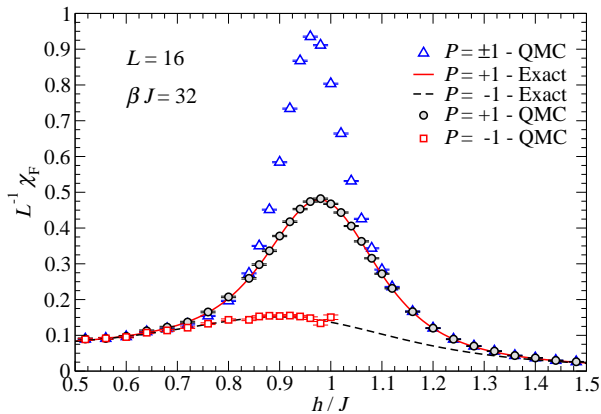}
   \end{center}
  \caption{(Color online) Fidelity susceptibility density for the 1$d$ TIM, as obtained from the exact solution (curves)
  and QMC simulations (symbols; see Sec.~\ref{sec:SSE_ChiF}). Results for both parity sectors
  $P=-1$ and $+1$ are shown (see Sec.~\ref{sec:parity}). QMC results indicated by $P=\pm 1$ (triangles) have been
  obtained by making no distinction between parity sectors.
  For $h/J \gtrsim1$, statistics are insufficient to estimate $\chi_{\rm F}$ in the $P=-1$ sector from QMC simulations
  since only the ground-state in the $P=+1$ sector is sampled at the low temperature considered here, $\beta J=32$.
  Data are for a system of size $L=16$ with periodic boundary conditions.}
  \label{fig:parity}
\end{figure}

\subsubsection{Parity quantum number}
\label{sec:parity}

An important issue concerning the TIM is the existence of a conserved quantum number, the parity
$P$. It is readily verified that the parity operator,
\begin{equation}
     {\mathcal P}= \prod_{i = 1}^{N} \sigma^{z}_{i}
  \label{eq:parity_op}
\end{equation}
(for a system with $N$ sites), commutes with the TIM Hamiltonian Eq.~(\ref{eq:tfim01}), $\left[ {\mathcal H},
{\mathcal P}\right]=0$, and therefore the parity $P=\pm 1$ is a good quantum number.

For finite systems, the ground-state of the TIM lies in the $P=+1$ sector, as shown by the following argument. It is convenient to work in the basis given by tensor products of the eigenvectors $\left\{ | \! \! \uparrow \rangle_x , | \! \! \downarrow \rangle_x \right\} $ of $\sigma_{i}^{x}$ at every site: $ \left\{ \ket{\phi_m} \right\} $ with, for instance, $\ket{\phi_m} = \ket{\! \! \uparrow \downarrow \downarrow
\uparrow \ldots}_x $. All off-diagonal matrix elements for the Hamiltonian Eq.~(\ref{eq:tfim01})
are non-positive in this basis and therefore the TIM on {\em finite lattices} satisfies the conditions for the Perron-Frobenius theorem to apply. According to this theorem, the coefficients of the system's ground-state (in its expansion in terms of the basis $\left\{ \ket{\phi_m} \right\}$, $\ket{\Psi_0} = \sum_m c_m \ket{\phi_m}$) must all have the same sign (say, $c_m \geq 0$). Consider the lowest-lying states in each parity sector $\ket{\Psi_{0}^{\pm}}$: they can be expanded as $\ket{\Psi_{0}^{\pm}} = \sum_m c_{m}^{\pm} \ket{\phi_{m}^{\pm}}$, with $\left\{ \ket{\phi_{m}^{\pm}} \right\} $ denoting the subset of elements in
$ \left\{ \ket{\phi_m} \right\} $ with fixed parity $P= \pm 1$. The argument proceeds by noticing that the parity operator
[Eq.~(\ref{eq:parity_op})] simply acts as a spin reversal operator upon the elements of the $\sigma^{x}$ basis:
namely, ${\cal P} \ket{\phi_{m}^{\pm}} = \ket{\psi_{m}^{\pm}}$, where $\ket{\psi_{m}^{\pm}}$ is obtained
from $\ket{\phi_{m}^{\pm}}$ by flipping all spins (for instance, ${\cal P} |  \! \!  \uparrow \downarrow \downarrow
\uparrow \ldots \rangle_x = | \! \! \downarrow \uparrow \uparrow \downarrow \ldots \rangle_x$).
Thus, the expression
\begin{equation*}
     {\mathcal P} \ket{\Psi_{0}^{\pm}} = \sum_m c_{m}^{\pm} {\mathcal P} \ket{\phi_{m}^{\pm}} =
     \sum_m c_{m}^{\pm} \ket{\psi_{m}^{\pm}} = \pm \ket{\Psi_{0}^{\pm}} ~,
\end{equation*}
is only consistent with the positiveness of the ground-state if $P=+1$ 
(one readily sees that the above relation can only be satisfied in the $P=-1$ sector if the coefficients
for the basis elements $\ket{\phi_{n}^{\pm}}$ and ${\mathcal P}\ket{\phi_{n}^{\pm}}=\ket{\psi_{n}^{\pm}}$
have opposite signs in the expansion for $\ket{\Psi_{0}^{-}}$).

The above discussion is directly relevant for our purposes here since, while expectation values for most physical
observables are the same for the lowest-lying states in both parity sectors, this turns out {\em not} to be the case
for $\chi_{\rm F}$ and $\chi_{\rm E}$. This is illustrated in Fig.~\ref{fig:parity}, where exact results for $\chi_{\rm F}$
(curves, see Ref.~\onlinecite{gu:08}) for the TIM on a chain with $L=16$ sites are shown for both parities, $P=+1$
and $P=-1$. Also shown are data obtained from a naive QMC implementation not discriminating between different
parity sectors (triangles in Fig.~\ref{fig:parity}): the disagreement is evident, specially in the neighborhood of the QCP
at $h_{\rm c} = 1$. While we expect this discrepancy to disappear in the thermodynamic limit (as suggested by
exact results for the $d=1$ TIM for increasingly larger systems; however, we have no general proof),
QMC simulations are obviously restricted to finite system sizes and it thus important to take the parity quantum
number into account.

Fortunately, the parity of a given SSE configuration $(\alpha, S_{n})$ can easily be determined within the here
adopted convention for the TIM [Eq.~(\ref{eq:tfim01})].\cite{convention} Indeed, the parity operator, defined in
Eq.~(\ref{eq:parity_op}), is diagonal in the $\sigma^{z}$ basis employed in SSE-QMC simulations
and the parity is then readily obtained as
\begin{equation*}
	P= \prod_{i = 1}^{N} \sigma^{z}_{i}\ket{\alpha (\tau = 0)}
\end{equation*}
(since the parity is a conserved number, it can be computed at any time-slice, in particular at $\tau = 0$). Both parity sectors are sampled due to the non-local updates in the QMC scheme (see Ref.~\onlinecite{syljuasen:02}). Therefore, our strategy consists in computing the estimators required for obtaining $\chi_{\rm F}$ and $\chi_{\rm E}$ (see Sec.~\ref{sec:sse}) for all SSE configurations $(\alpha, S_{n})$ and storing the results in different variables according to the parity of the state $\ket{\alpha (\tau = 0)}$. Data obtained in this way for the $d=1$ TIM are shown in Fig.~\ref{fig:parity} and perfectly agree with the exact results.
Finally, since for finite systems the ground-state has $P=+1$, all results discussed in the present work have been obtained for this parity sector (with the exception of the indicated ones in Fig.~\ref{fig:parity}).

\subsection{Simulation Details}
\label{sec:pre_qmc}

\begin{figure}
  \begin{center}
    \includegraphics*[width=0.3\textwidth,angle=270]{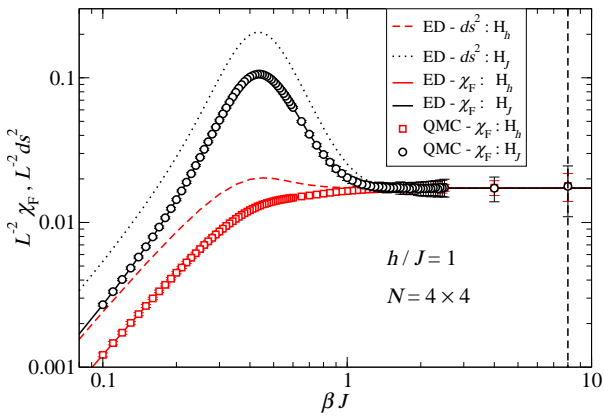}
   \end{center}
  \caption{(Color online) Finite-$T$ fidelity susceptibility $L^{-2}\chi_{\rm F}(g,\beta)$
  [Eqs.~(\ref{eq:ChiF_T01},\ref{eq:ChiF_T02})] and Bures metric $L^{-2}ds^{2}(g,\beta)$
  [Eq.~(\ref{eq:Bures02})] as a function of the inverse temperature $\beta=T^{-1}$
  for the TIM on the square lattice ($N=4 \times 4$ sites cluster with periodic boundary
  conditions; restriction to the $P=+1$ parity sector, see Sec.~\ref{sec:parity}).
  Both $H_{J}$ and $hH_{h}$ [see Eq.~(\ref{eq:tfim01})] have been considered as the
  ``driving term" $H_1$ in the definitions for $\chi_{\rm F}(g,\beta)$ and $ds^{2}(g,\beta)$:
  we see that different choices yield the same $T=0$ result. The vertical dashed line marks
  the value $\beta=2L$ set in obtaining the QMC data displayed in Figs.~\ref{fig:parity} and
  \ref{fig:chiF}: while for the coupling considered here ($h/J=1$) the system is deep into
  the ferromagnetic phase and convergence to ground-state expectation values is achieved for
  smaller $\beta$, the more stringent condition $\beta=2L$ is necessary closer to the QCP.
  }
  \label{fig:partition}
\end{figure}

The computation of $\chi_{\rm F}(g,\beta)$ and $\chi_{\rm E}(g,\beta)$ requires only small changes to an
existing SSE code: estimators for both quantities [Eqs.~(\ref{eq:ChiF_QMC}) and (\ref{eq:d2E0_QMC}),
respectively] are simply computed by analyzing the operator strings [Eq.~(\ref{eq:Sn})], a task ideally
carried out while performing diagonal updates for the SSE configurations $(\alpha, S_{n})$.\cite{syljuasen:02}
Our code is based on the ALPS\cite{albuquerque:07} librairies implementation of SSE QMC.\cite{alet:05a}
The main modifications of the original codes are independent from measurements for $\chi_{\rm F}$ and $\chi_{\rm E}$
and are specific to the TIM [as defined by Eq.~(\ref{eq:tfim01})] studied in the present work.
They involve changes in the processes that are allowed when performing off-diagonal updates\cite{updates} and the
computation of the parity quantum number (see Sec.~\ref{sec:parity}).

Calculating $\chi_{\rm F}$ can be computationally demanding due to the fact that the number of operations
required for obtaining the histogram $N_{gH_1}(m)$ [see Eq.~(\ref{eq:ChiF_QMC})] scales quadratically
with the total number of $g H_1$ operators in the string $S_{n}$. The situation can be ameliorated by
a judicious bipartition of the system's Hamiltonian into $H_0$ and $H_1$ [Eq.~(\ref{eq:hamiltonian})]. 

Indeed, there is freedom to consider either $H_{J}$ or $hH_{h}$ (or, in general, any combination
of these) appearing in Eq.~(\ref{eq:tfim01}) as the ``driving term" $H_1$ in Eq.~(\ref{eq:ChiF_pert}):
since $\bra{\Psi_{n} (h)}  (H_{J} + hH_{h})  \ket{\Psi_{0} (h)} = 0$ for $n\neq 0$, we readily conclude from
Eq.~(\ref{eq:ChiF_pert}) that $\chif^{H_{J}} = h^2 \chif^{hH_{h}}$ (superscripts indicate the term assigned
to $H_1$) and therefore different bipartition choices lead to the same zero-temperature results for
$\chi_{\rm F}$, apart from a trivial multiplicative factor. Although this is no longer true for the finite-$T$
generalizations of Eqs.~(\ref{eq:Bures02}) and (\ref{eq:ChiF_T02}),\cite{finiteT} the equivalence between
results obtained from different partitions is recovered in the limit $T \rightarrow 0$, as shown in Fig.~\ref{fig:partition}
for the TIM on the square lattice (data obtained from QMC simulations for a $4 \times 4$ cluster and
$h/J=1$).

We may thus explore the fact that different terms in the Hamiltonian dominate in different regions of
the phase diagram in order to reduce computational cost. Specifically, for the case of the TIM
considered here [Eq.~(\ref{eq:tfim01})] it is more efficient to compute $N_{g H_1}(m)$ in the high-field limit
if we set $H_1=H_{J}$, since the number of such operators in the strings $S_{n}$ will be smaller than that
of ${H_h}$ operators in this limit. In practice, we find that, for field magnitudes close to the QCP
for the TIM on the square lattice, setting $H_1=H_{J}$ is the most efficient choice.

Following this strategy, we have simulated the TIM on the square lattice by considering clusters
with linear size $L$ and periodic boundary conditions (PBC), and are able to reach $L=28$ when
computing $\chif$. On the other hand, computing  $\chi_{\rm E}$ requires much lesser numerical
effort and we are able to reach $L=48$. For both quantities, we find that if we set the inverse
temperature $\beta=2L$ both $\chif$ and $\chi_{\rm E}$
reach their ground-state expectation values. This is illustrated in Fig.~\ref{fig:partition} for the $L=4$
cluster. We have also performed a few simulations setting $\beta=4L$ in order to confirm that
convergence has indeed been achieved, at least within error bars, for $\beta=2L$.

\begin{figure}
  \begin{center}
    \includegraphics*[width=0.3\textwidth,angle=270]{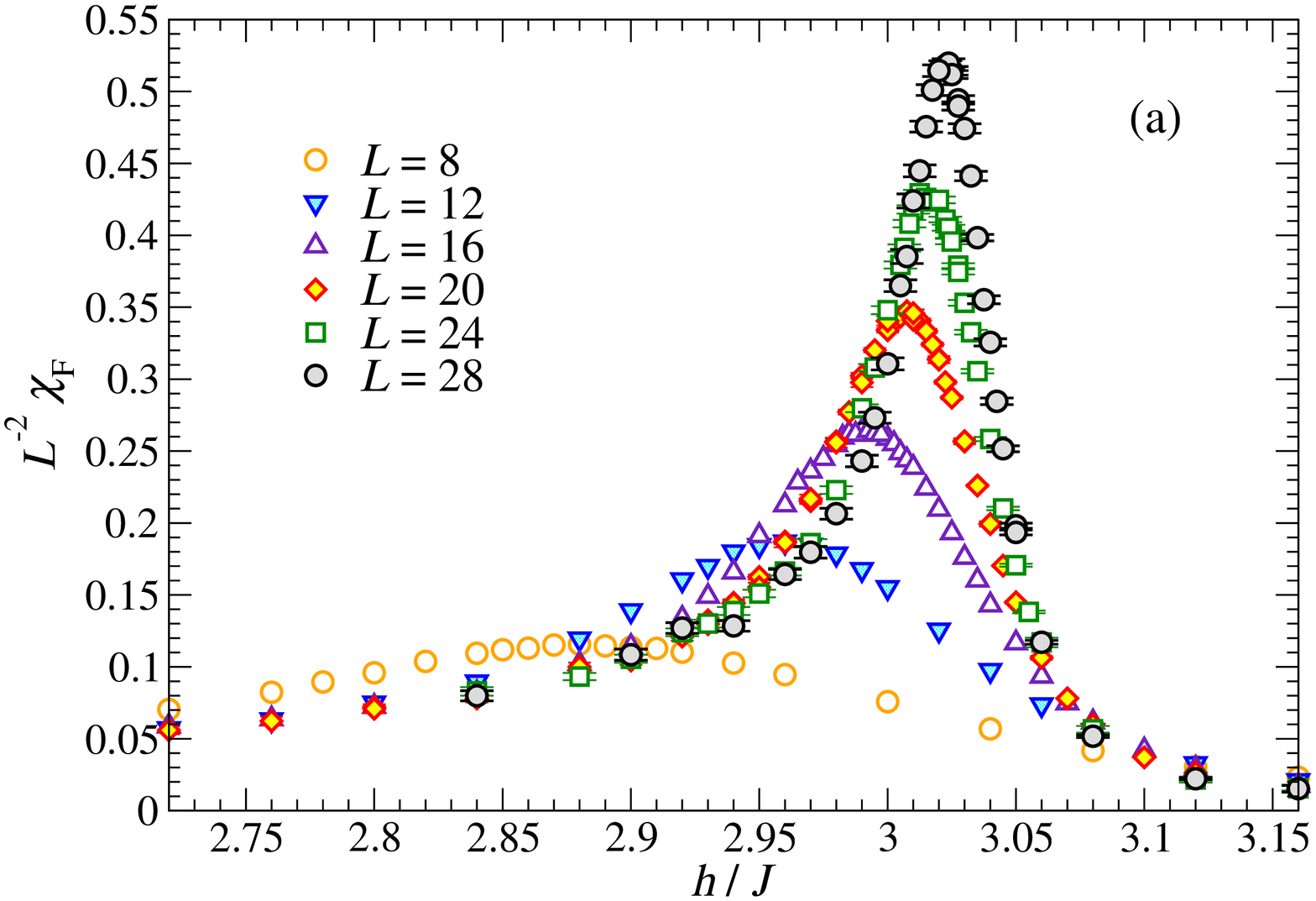}
    
    \vspace{0.2cm}
   
    \includegraphics*[width=0.3\textwidth,angle=270]{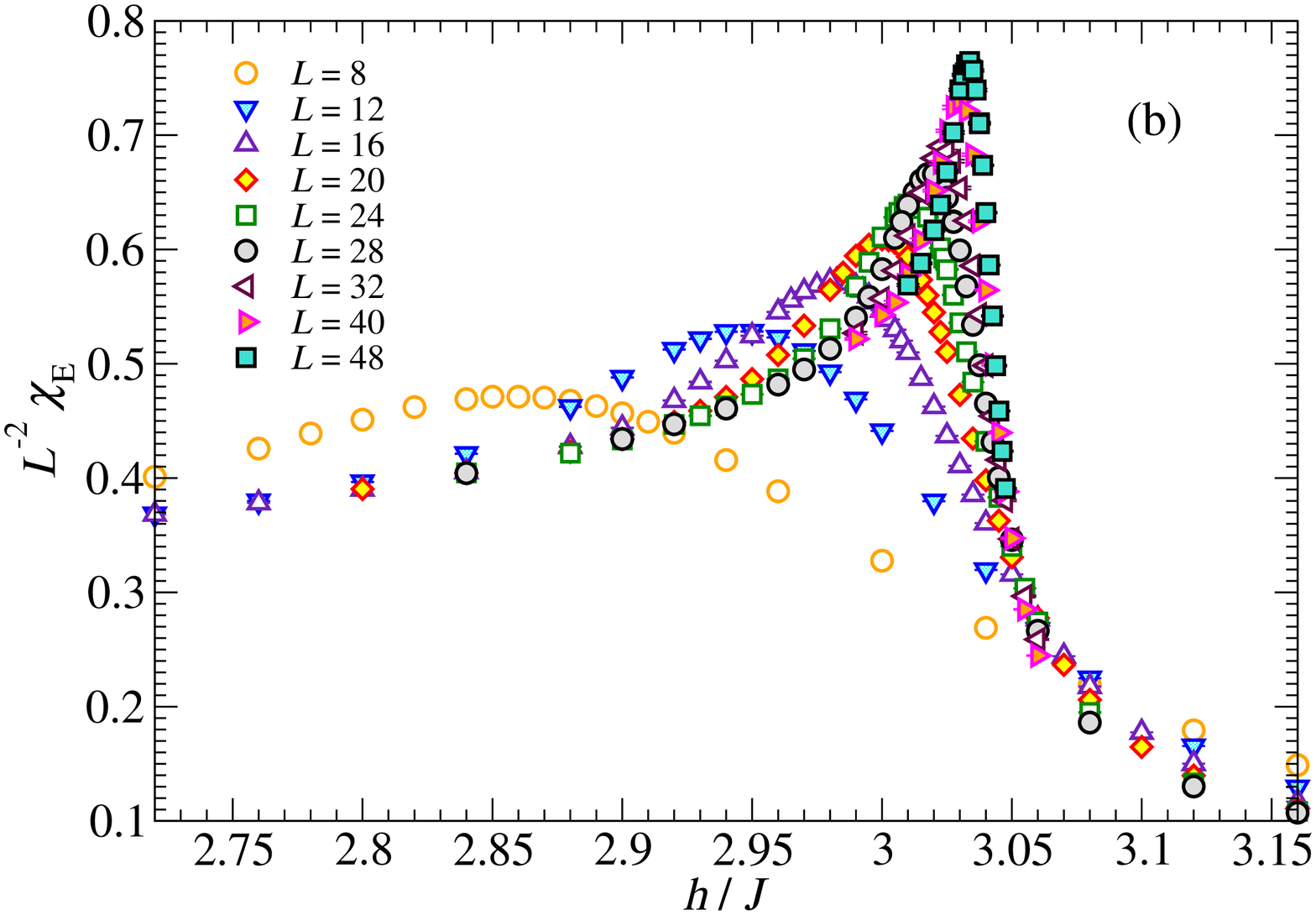} 
   \end{center}
  \caption{(Color online) (a) Fidelity susceptibility density $L^{-2}\chi_{\rm F}$ and (b) ground-state energy's
  second derivative per site $L^{-2}{\partial^{2} E_{\rm 0}(g)}/{\partial g^{2}}=-L^{-2}\chi_{\rm E}(g)$
  for the TIM on the square lattice, as a function of $h/J$ and for indicated system sizes $L$
  (temperatures are set to $\beta=2L$).
  Data have been obtained by applying the SSE QMC procedure detailed in
  Sec.~\ref{sec:sse}.}
  \label{fig:chiF}
\end{figure}

\subsection{Results}
\label{sec:results}

Our QMC data for $L^{-2}\chi_{\rm F}$ and $L^{-2}\chi_{\rm E}$ for the TIM on the square lattice are shown
in Fig.~\ref{fig:chiF} for various system sizes $L$. The presence of peaks in the curves for both quantities is evident:
they become more pronounced for increasing $L$ and their positions seemingly converge toward the
estimate $h_{\rm c}=3.04438(2)$ for the QCP found in Ref.~\onlinecite{bloete:02} (see below). Furthermore,
we notice that $\chi_{\rm E}$ displays less pronounced peaks than $\chi_{\rm F}$, as expected from our discussion
in Sec.~\ref{sec:scaling}. A quantitative data analysis is explained in what follows.

\begin{figure}
  \begin{center}
     \includegraphics*[width=0.28\textwidth,angle=270]{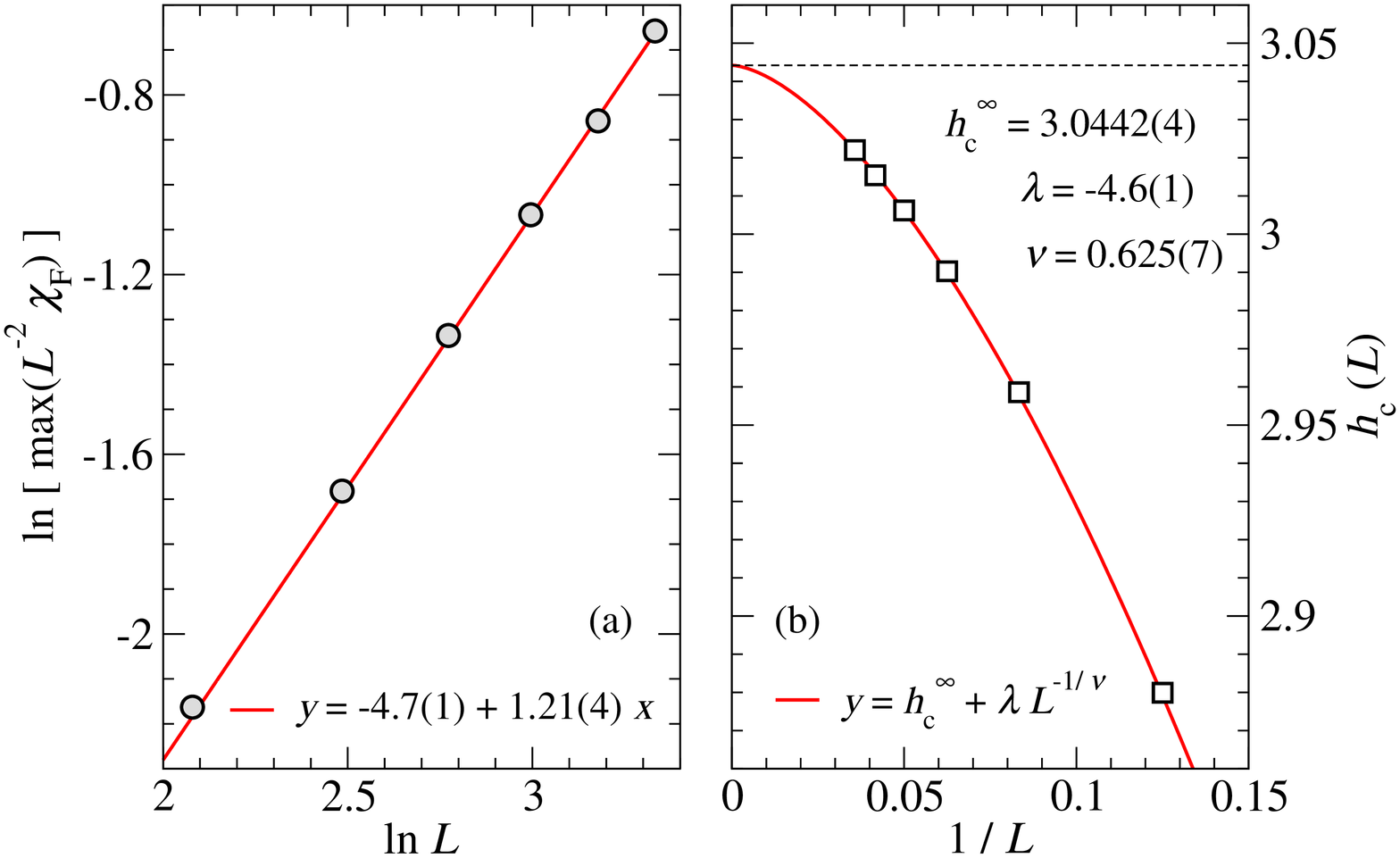}
    
    \vspace{0.2cm}
   
    \includegraphics*[width=0.28\textwidth,angle=270]{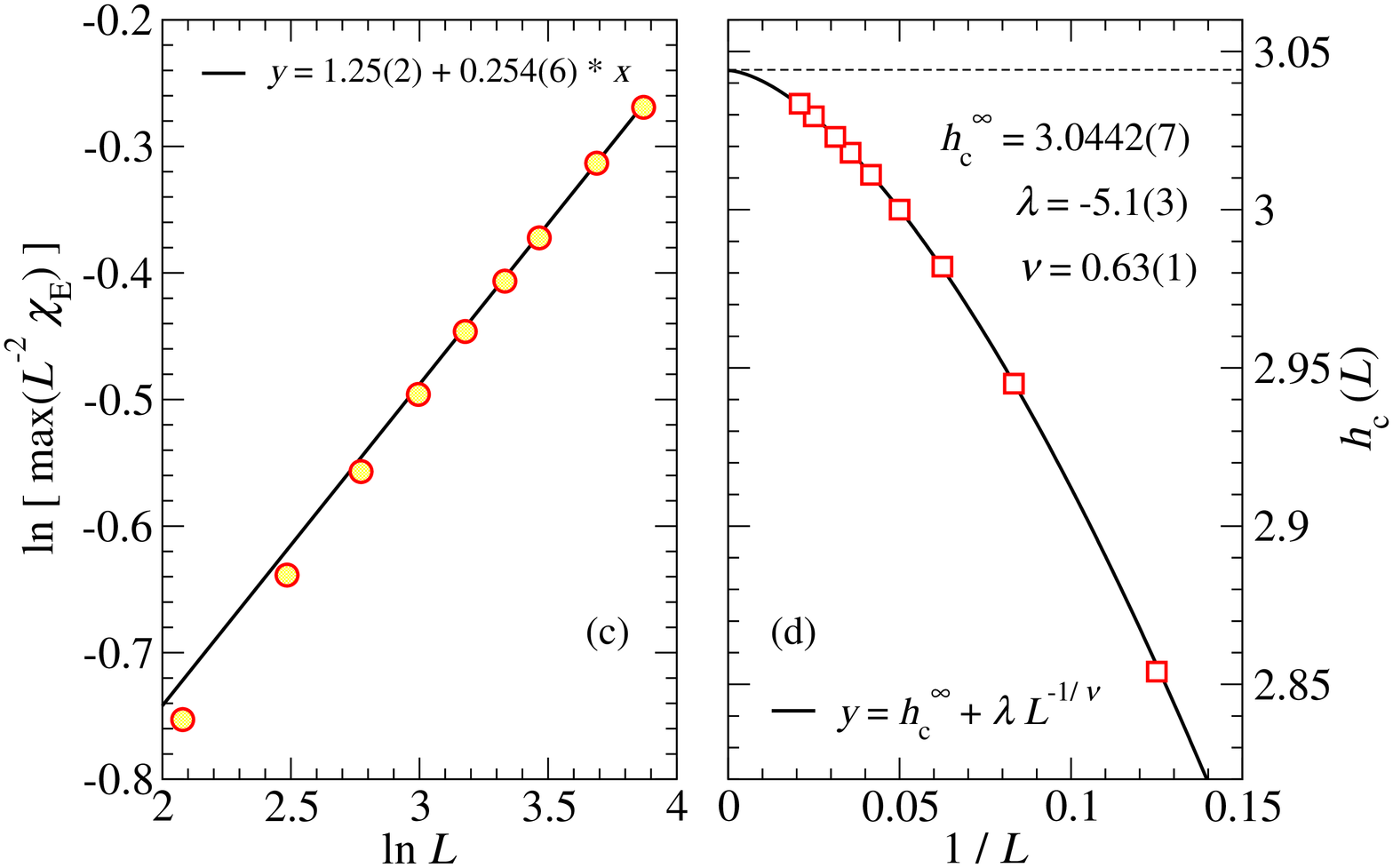}
   \end{center}
  \caption{(Color online) Finite size scaling analysis for the location and height of the
  peaks in $\chi_{\rm F}$ [(a) and (b)] and $\chi_{\rm E}$ [(c) and (d)], obtained from
  the QMC data shown in Fig.~\ref{fig:chiF}. In panels (a) and (c), the logarithm of the
  maxima in $L^{-2}\chi_{\rm F}$ and $L^{-2}\chi_{\rm E}$, respectively, are plotted
  as function of $\ln L$. Linear regression (lines) is applied to the three rightmost
  data-points in each case, yielding the estimates (a) $\nu=0.623(8)$ and (c) $\nu=0.615(1)$
  for the correlation length's critical exponent. In (b) and (d), the peaks' location $h_{\rm c}(L)$ for,
  respectively, $L^{-2}\chi_{\rm F}$ and $L^{-2}\chi_{\rm E}$ are plotted against
  inverse system size $1/L$. Fits (curves) for these results by using Eq.~(\ref{eq:hc_L})
  yield the estimates: (b) $h_{\rm c}^{\infty}=3.0442(4)$ and $\nu=0.625(7)$ and
  (d) $h_{\rm c}^{\infty}=3.0442(7)$ and $\nu=0.63(1)$ (the extrapolated values
  $h_{\rm c}^{\infty}$ are indicated by the horizontal dashed lines). See main text for details.
  }
  \label{fig:peaks}
\end{figure}

We start by determining the peaks positions and heights for both $\chi_{\rm F}$ and $\chi_{\rm E}$
from the raw data displayed in Fig.~\ref{fig:chiF}. The so obtained results are shown in Fig.~\ref{fig:peaks}.
From the scaling relations derived in Sec.~\ref{sec:scaling}, $L^{-d}\chi_{\rm F} \sim L^{\frac{2}{\nu} - d}$
and $L^{-d}\chi_{\rm E} \sim L^{\frac{2}{\nu} - (d+z)}$ [Eqs.~(\ref{eq:dim_chiF_04}) and (\ref{eq:dim_d2E0_03});
$d=2$ and $z=1$], we expect a linear dependence for the logarithm of the peaks' height on $\ln L$. This
is confirmed by the results shown in Figs.~\ref{fig:peaks}(a) and (c). By applying linear regression to the points
associated to the three largest values of $L$ in each plot we obtain our first estimates for correlation length's
exponent: $\nu=0.623(8)$ [$\chi_{\rm F}$, Fig.~\ref{fig:peaks}(a)] and $\nu=0.615(1)$ [$\chi_{\rm E}$,
Fig.~\ref{fig:peaks}(c)]. While the former estimate is in good agreement with the result for the
universality class of the three-dimensional classical Ising model [$\nu=0.6301(8)$, Ref.~\onlinecite{bloete:95}], the latter
clearly underestimates $\nu$. This is likely to be explained by the weak divergence displayed by
$\chi_{\rm E}$, implying that regular sub-leading corrections are important in accounting for the
behavior in system sizes as the ones considered here: indeed we notice that
the data points corresponding to the smallest system sizes clearly deviate from the linear fit
obtained for the points for the three largest $L$ in Fig.~\ref{fig:peaks}(c).

In Figs.~\ref{fig:peaks}(b) and (d) we plot the peaks' location versus inverse system size $1/L$
for $\chi_{\rm F}$ and $\chi_{\rm E}$, respectively. We expect (see for instance the related discussion
in Ref.~\onlinecite{venuti:08b})
the following expression to hold for the scaling
of the peak positions for $h_{\rm c} (L)$ with system size $L$
\begin{equation}
    h_{\rm c} (L) = h_{\rm c}^{\infty} + \frac{\lambda}{L^{1/\nu}}~,
  \label{eq:hc_L}
\end{equation}
where $h_{\rm c}^{\infty}$ is the result for $L \rightarrow \infty$. Data fits give the following
estimates: $h_{\rm c}^{\infty}=3.0442(4)$ and $\nu=0.625(7)$ [$\chi_{\rm F}$, Fig.~\ref{fig:peaks}(b)]
and $h_{\rm c}^{\infty}=3.0442(7)$ and $\nu=0.63(1)$ [$\chi_{\rm E}$, Fig.~\ref{fig:peaks}(d)].
We remark that our estimates for the location of the QCP are in very good agreement with the result
from Ref.~\onlinecite{bloete:02} and, although quality is lesser in this case, our results for $\nu$
are consistent with the value $\nu=0.6301(8)$ found in Ref.~\onlinecite{bloete:95}.

\begin{figure}
  \begin{center}
   \includegraphics*[width=0.3\textwidth,angle=270]{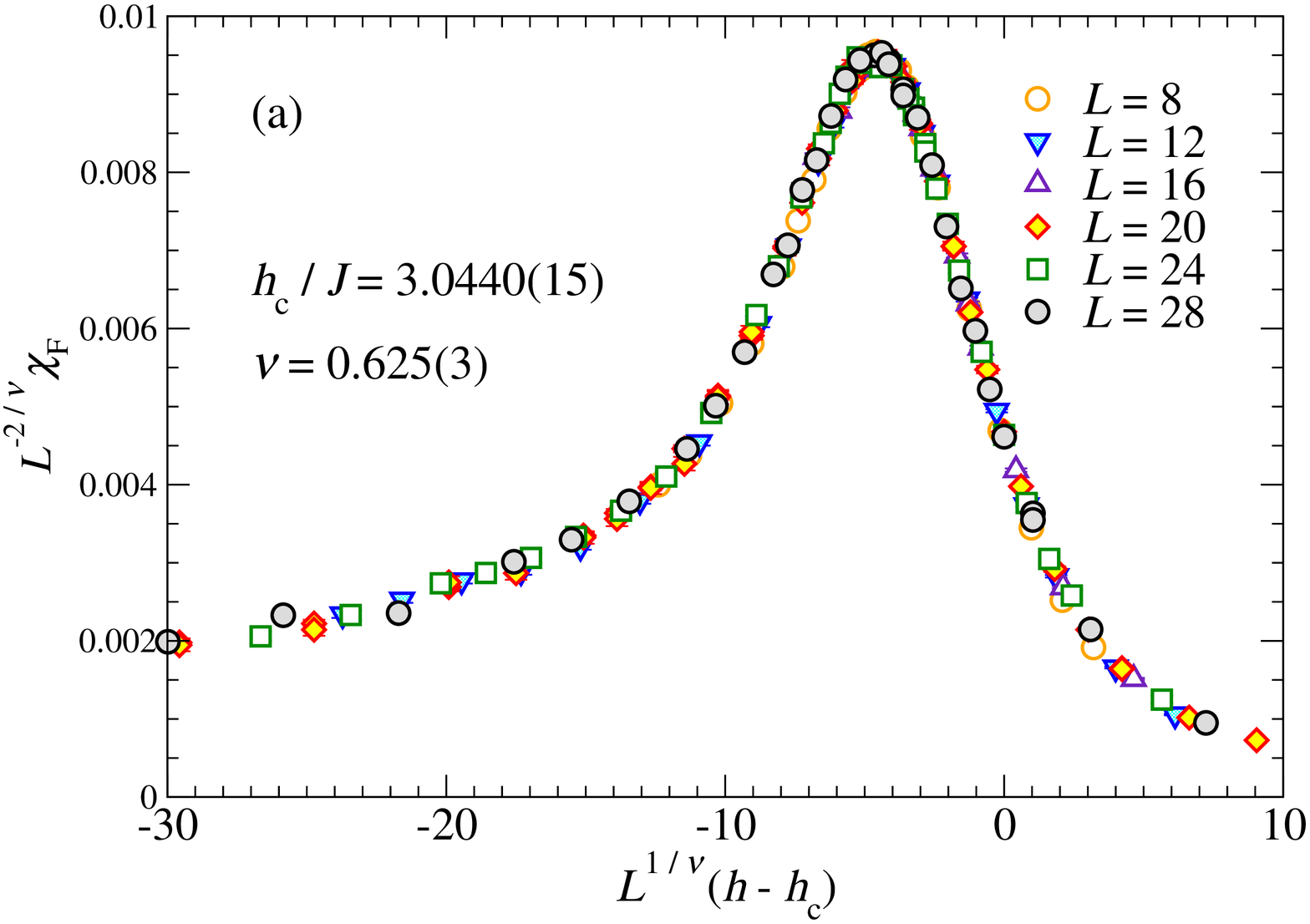}

    \vspace{0.2cm}
   
    \includegraphics*[width=0.3\textwidth,angle=270]{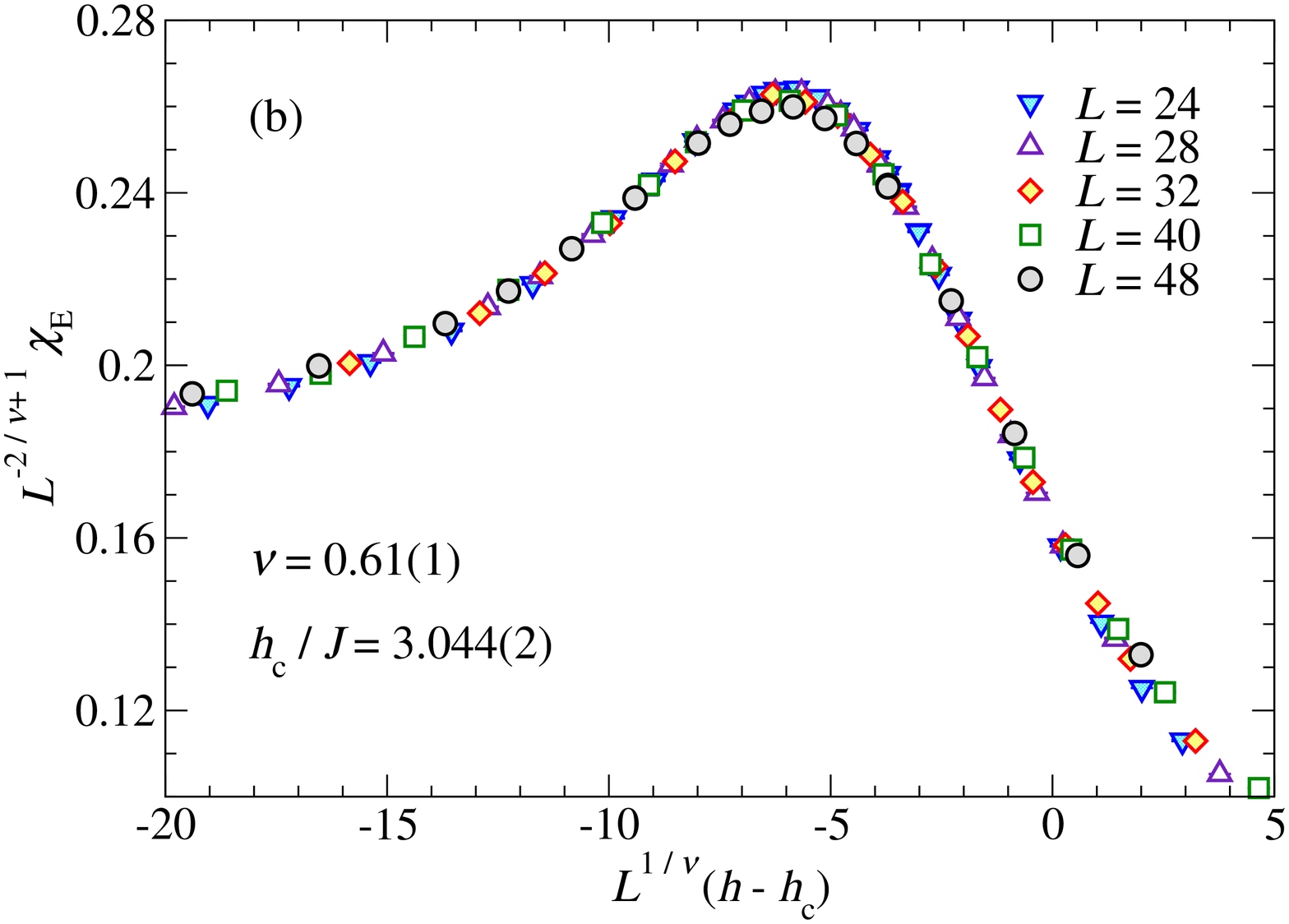}
   \end{center}
  \caption{(Color online) (a) Data collapse for the QMC results for $L^{-2}\chi_{\rm F}$ (a)
  and $L^{-2}\chi_{\rm E}$ (b) for the TIM on the square lattice and indicated system
  sizes. Data collapse is achieved
  for: (a) $h_{\rm c}=3.0440(15)$ and $\nu=0.625(3)$ and (b) $h_{\rm c}=3.044(3)$ and $\nu=0.61(1)$.
  In panel (b), data for the smallest system sizes are discarded (see main text).}
  \label{fig:chiF_collapse}
\end{figure}

Finally, from the finite size scaling analysis performed in Sec.~\ref{sec:scaling} we expect the following
relation to describe the behavior of $\chi_{\rm F}$ on finite systems in the neighborhood of the QCP
\begin{equation}
   L^{-d} \chi_{\rm F} (h, L) = L^{\frac{2}{\nu} - d} f_{\chi_{\rm F}}\left( L^{1/\nu} | h - h_{\rm c} | \right)~,
  \label{eq:ChiF_collapse}
\end{equation}
and similarly for $\chi_{\rm E}$
\begin{equation}
   L^{-d} \chi_{\rm E} (h, L) = L^{\frac{2}{\nu} - (d+z)} f_{\chi_{\rm E}}\left( L^{1/\nu} | h - h_{\rm c} | \right)~.
  \label{eq:ChiE_collapse}
\end{equation}
In the above expressions $f_{\chi_{\rm F}}$ and $f_{\chi_{\rm E}}$ are homogeneous functions,
{\em a priori} unknown. Estimates for critical parameters can thus be obtained by plotting
$L^{-\frac{2}{\nu}}\chi_{\rm F}$ and $L^{-\frac{2}{\nu} + z}\chi_{\rm E}$ versus $L^{1/\nu} | h - h_{\rm c} |$
and adjusting the values of $h_{\rm c}$ and $\nu$ until data collapse is achieved. The so obtained data
collapse plots are displayed in Fig.~\ref{fig:chiF_collapse}, from which we get the following estimates:
$h_{\rm c}=3.0440(15)$ and $\nu=0.625(3)$ [$\chi_{\rm F}$, Fig.~\ref{fig:chiF_collapse}(a)],
and $h_{\rm c}=3.044(2)$ and $\nu=0.61(1)$ [$\chi_{\rm E}$, Fig.~\ref{fig:chiF_collapse}(b)].
All of these values are in agreement with published results [$h_{\rm c}=3.04438(2)$ from Ref.~\onlinecite{bloete:02}
and $\nu=0.6301(8)$, Ref.~\onlinecite{bloete:95}], but we remark the slightly lesser quality of the data collapse achieved for
$\chi_{\rm E}$. Indeed, data obtained from the smallest system sizes in Fig.~\ref{fig:chiF}(b)
fail to collapse onto the curve for the largest $L$ in Fig.~\ref{fig:chiF_collapse}(b) and have not
been taken into account when performing the analysis. Again, we believe that this is explained by
the weakly divergent behavior of $\chi_{\rm E}$.

\section{Discussion and conclusions}
\label{sec:conclusions}

In summary, we have investigated the scaling properties of the fidelity susceptibility $\chi_{\rm F}$
in the quantum critical regime. Large scale quantum Monte Carlo simulations for the
transverse-field Ising model on the square lattice, performed by using the scheme
introduced in Ref.~\onlinecite{schwandt:09}, confirm the validity of the derived scaling relations.
Additionally, we also investigate the scaling behavior of the ground-state energy's second
derivative, ${\partial^{2} E_{\rm 0}(g)}/{\partial g^{2}} = -\chi_{\rm E}(g)$, a quantity closely
related to $\chi_{\rm F}$.

We would like to highlight the fact that the novel QMC scheme for computing $\chi_{\rm F}$
presented in Ref.~\onlinecite{schwandt:09} and discussed in detail in the present work
opens several research possibilities within the so-called fidelity approach to quantum
critical phenomena. Indeed, investigations in this field have been so far, to a large extent, restricted
to one-dimensional systems, while our QMC scheme allows for the study of $\chi_{\rm F}$
for a large class of sign-problem-free models in arbitrary dimensions. Furthermore, we stress that
the required modifications in a pre-existing SSE code are minimal and that, even though SSE is
particularly well suited for the task, it is also likely that measurements for $\chifg$ can be implemented
within other QMC flavors, such as the loop algorithm.\cite{evertz:03} Another potentially interesting
possibility opened by the QMC method considered here involves the study of the finite-$T$ properties
of $\chi_{\rm F}$ (which scales as the Bures metric, as shown in Sec.~\ref{sec:finiteT}), along the lines of
Ref.~\onlinecite{zanardi:07}.

A second point worth to emphasize is the particularly simple scaling relations for $\chi_{\rm F}$
derived in Sec.~\ref{sec:scaling}, expressed {\em solely} in terms of the correlation length's
critical exponent $\nu$ and that considerably extends the result obtained by Campos Venuti
and Zanardi.\cite{venuti:07} Perhaps even more importantly, our scaling analysis does not rely
on novel concepts such as ``quantum adiabatic dimension" recently advocated by Gu and
coworkers.\cite{gu:09a,gu:09b} Also, and to the best of our knowledge, our Eq.~(\ref{eq:dim_chiF_04})
is consistent with several results for the scaling behavior of $\chi_{\rm F}$ close to second order QCPs
presented in the literature, including those compiled in Table I of the review Ref.~\onlinecite{gu:08}
(note that some results from the litterature quoted in this table mistake $\nu$ for $1/\nu$).

Additionally, we also obtain a scaling relation for $\chi_{\rm E}$,
something that allows us to address the important point of which of the two quantities,
$\chi_{\rm F}$ or $\chi_{\rm E}$, is better suited in detecting quantum phase transitions.
This question is particularly important from the perspective opened by the QMC SSE scheme:
indeed, as we discuss in detail in Sec.~\ref{sec:sse}, the computational cost for calculating
$\chi_{\rm E}$ can be orders of magnitude smaller than the one required in obtaining $\chi_{\rm F}$,
something in favor of the former as a better indicator of quantum criticality, from a practical perspective. However, our
scaling analysis shows (Sec.~\ref{sec:scaling}) that $\chi_{\rm E}$ exhibits a weaker divergence
(by a factor of $z$ in the exponent) than $\chi_{\rm F}$, meaning that it might be necessary to
take into account non-divergent sub-leading corrections when performing finite size scaling
analysis for $\chi_{\rm E}$. Furthermore, there may even
be situations where only $\chi_{\rm F}$ diverges: according to the scaling theory, this
happens whenever $2/(d+z) < \nu < 2/d$. As a concrete example, we mention
the case of the QCPs for the CaVO system analyzed in Ref.~\onlinecite{schwandt:09}, that are
preferably detected as a divergence in $\chif$ rather than as a cusp in $\chi_{\rm E}$.
 
That is, the question of which among $\chi_{\rm F}$ and $\chi_{\rm E}$ is best tailored to detect
an unknown QCP depends on its (possibly unknown) universality class and on practical matters
such as the system sizes that can be reached within SSE QMC. From a practical point of view,
a possible strategy consists in evaluating $\chi_{\rm E}$ on system of up to intermediate sizes
(where simulations are not too demanding) and search for the presence of peaks hinting at
a singularity or a cusp in the thermodynamic limit. In the affirmative case, simulations for larger
systems sizes may be performed in order to confirm the occurrence of singular behavior for this quantity.
If this is not the case, one should measure $\chifg$ for intermediate sizes and check on whether
a singularity is more apparent.

More specifically, we compare now our results for the TIM on the square lattice to those obtained,
through means of exact diagonalizations, by Yu and coworkers.\cite{yu:09}
Yu {\em et al.}~have been able to study $\chi_{\rm F}$ and $\chi_{\rm E}$ by considering
clusters comprising up to $20$ sites and have arrived to the following estimates of
critical parameters: $h_{\rm c}=2.95(1)$ and $\nu \simeq 1.40$. On the other hand, by
resorting on the SSE QMC method discussed in Sec.~\ref{sec:sse} and on the scaling relations
in Sec.~\ref{sec:scaling}, we are able to compute $\chi_{\rm F}$ for systems with up to $N=28 \times 28$
sites and $\chi_{\rm E}$ for systems with up to $N=48 \times 48$, arriving at the estimates
(from the data collapse for $\chi_{\rm F}$ performed in Sec.~\ref{sec:results}): $h_{\rm c}=3.0440(15)$
and $\nu=0.625(3)$. Our estimate for the location of the QCP clearly compares much better with results
from conventional approaches [$h_{\rm c}=3.04438(2)$ from Ref.~\onlinecite{bloete:02}] than the one found
in Ref.~\onlinecite{yu:09}. And, even more importantly, while our result for $\nu$ is in good
agreement with the known result for the universality class of the classical Ising model in $d=3$
[$\nu=0.6301(8)$, Ref.~\onlinecite{bloete:95}], the value for $\nu$ quoted in Ref.~\onlinecite{yu:09}
considerably deviates from it. Again, we suspect that the value for $\nu$ is incorrectly presented as the value for $1/\nu$.

The fact that the analysis employed in Ref.~\onlinecite{yu:09} fails to obtain critical parameters in
agreement with the ones from conventional approaches highlights the importance of the two main results
presented here. First, our SSE QMC allows for the computation of $\chi_{\rm F}$ and $\chi_{\rm E}$ for much
larger systems than possible within exact diagonalizations, enormously improving the quality of finite-size
scaling analysis (we remark that results for clusters comprising less than $N=8 \times 8$ sites are not
even taken into account in the data collapse performed in Sec.~\ref{sec:results}). Second, the scaling
relations derived in Sec.~\ref{sec:scaling} extends previous results\cite{venuti:07} and
expresses the scaling dimensions for both $\chi_{\rm F}$ and $\chi_{\rm E}$ in terms of the correlation
length exponent. This has the advantage that the exponents obtained for $\chi_{\rm F}$
and $\chi_{\rm E}$ can be directly compared to established results for a given universality class,
allowing us to decide on the validity of the approach.

We also remark that the scaling relations derived here are in agreement with the ones recently derived
in the field of quantum quenches.\cite{grandi:10,barankov:09,grandi:09} In this context, the fidelity (and
its susceptibility) governs the probability for the system to transit to an excited state after a sudden change
of the coupling constant $g$ away from the critical point $g_c$. This expands the range of applicability of
the concept of fidelity susceptibility beyond the fidelity approach to quantum phase transitions.\cite{gu:08}
We might therefore expect that the QMC method presented here, or an adaptation thereof, is also applicable
in this context.  


\begin{acknowledgments}
We acknowledge fruitful exchanges with O.~Motrunich, A.~Polkovnikov, G.~Roux, D.~Schwandt and L.~Campos
Venuti.
Calculations were performed using the SSE code~\cite{alet:05a} of the ALPS libraries.\cite{albuquerque:07}
We thank CALMIP for allocation of CPU time. 
This work is supported by the French ANR program ANR-08-JCJC-0056-01.
\end{acknowledgments}


\appendix
\section{Analytical approximation of an integral}
\label{appendix}
In this section, we derive useful approximate analytical expressions for Eq.~(\ref{eq:integral}), 
\begin{equation*}
A(m,n)=\frac{(n-1)!}{m!(n-m-2)!}\int_{0}^{1/2}\tau^{m+1}(1-\tau)^{n-m-2}\,d\tau~.
\end{equation*}
First, we note that $A(m,n)$ can be written in a more symmetric form
\begin{equation*}
A(m,n)=\frac{m+1}{n}f(m+1,n-m-2),
\end{equation*}
where
\begin{equation*}
f(p,q)=\frac{(p+q+1)!}{p!q!}\int_{0}^{1/2}\tau^{p}(1-\tau)^{q}\,d\tau.
\end{equation*}
We now concentrate on finding efficient analytical approximations for
$f(p,q)$, in the limit where $r=p+q+1$ is large. Ultimately, we can
apply these estimates to our practical case, corresponding to $p=m+1$,
$q=n-m-2$, $r=n$.

$f(p,q)$ can be alternatively written
\begin{equation*}
f(p,q)=\frac{\int_{0}^{1/2}\tau^{p}(1-\tau)^{q}\,d\tau}
{\int_{0}^{1}\tau^{p}(1-\tau)^{q}\,d\tau}.
\end{equation*}
After the change of variable $\tau=\frac 12\left(1-\frac{t}{\sqrt{p+q}}\right)$, we obtain
\begin{equation*}
f(p,q)=\frac{\int_{0}^{\sqrt{p+q}}\left(1-\frac{t^2}{p+q}\right)^{\frac{p+q}{2}}
\left(\frac{1+\frac{t}{\sqrt{p+q}}}{1-\frac{t}{\sqrt{p+q}}}\right)^{\frac{q-p}{2}}\,dt}
{\int_{-\sqrt{p+q}}^{\sqrt{p+q}}\left(1-\frac{t^2}{p+q}\right)^{\frac{p+q}{2}}
\left(\frac{1+\frac{t}{\sqrt{p+q}}}{1-\frac{t}{\sqrt{p+q}}}\right)^{\frac{q-p}{2}}\,dt}.
\end{equation*}
As a first approximation in the limit $r\to\infty$, keeping fixed the ratio
\begin{equation}
X_1(p,q)=\frac{q-p}{\sqrt{r}},\label{X1}
\end{equation}
we obtain
\begin{equation*}
f(p,q)\approx\frac{\int_{0}^{\infty}{\rm e}^{-\frac{t^2}{2}+X_1(p,q)t}\,dt}
{\int_{-\infty}^{\infty}{\rm e}^{-\frac{t^2}{2}+X_1(p,q)t}\,dt},
\end{equation*}
which simplifies into 
\begin{equation}
f(p,q)\approx G[X_1(p,q)],\label{order1}
\end{equation}
where
\begin{eqnarray*}
G(x)&=&\frac{1}{\sqrt{2\pi}}\int_{-\infty}^x{\rm e}^{-y^2/2}\,dy,\\
&=&\frac 12\left[1+{\rm Erf}\left(\frac{x}{\sqrt 2}\right)\right].
\end{eqnarray*}

In order to obtain more systematic and more rigorous approximate
analytic expressions for $f(p,q)$, we quite generally introduce
$X(p,q)$ such that
\begin{equation}
f(p,q)\equiv G[X(p,q)].\label{defeq}
\end{equation}
Since for $p\to +\infty$ (at fixed $q$), one has $f(p,q)\to 0$, we see
that $X(p,q)\to-\infty$ in this limit. Moreover, the symmetry $f(p,q)=1-f(q,p)$
implies that $X(p,q)=-X(q,p)$.

\begin{figure}
  \begin{center}
    \includegraphics*[width=0.3\textwidth,angle=270]{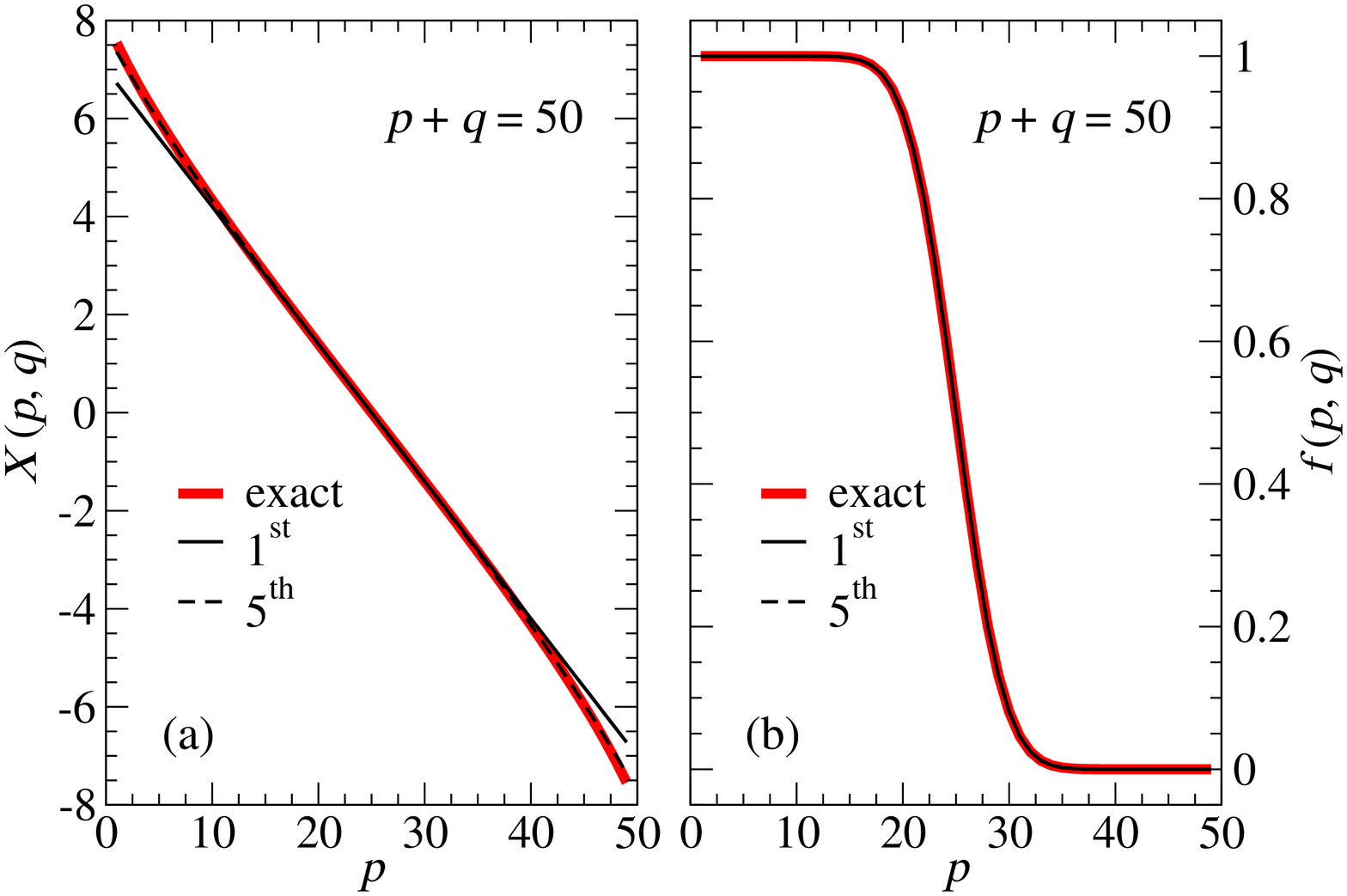}
    \vspace{0.2cm}
    \includegraphics*[width=0.17\textwidth,angle=270]{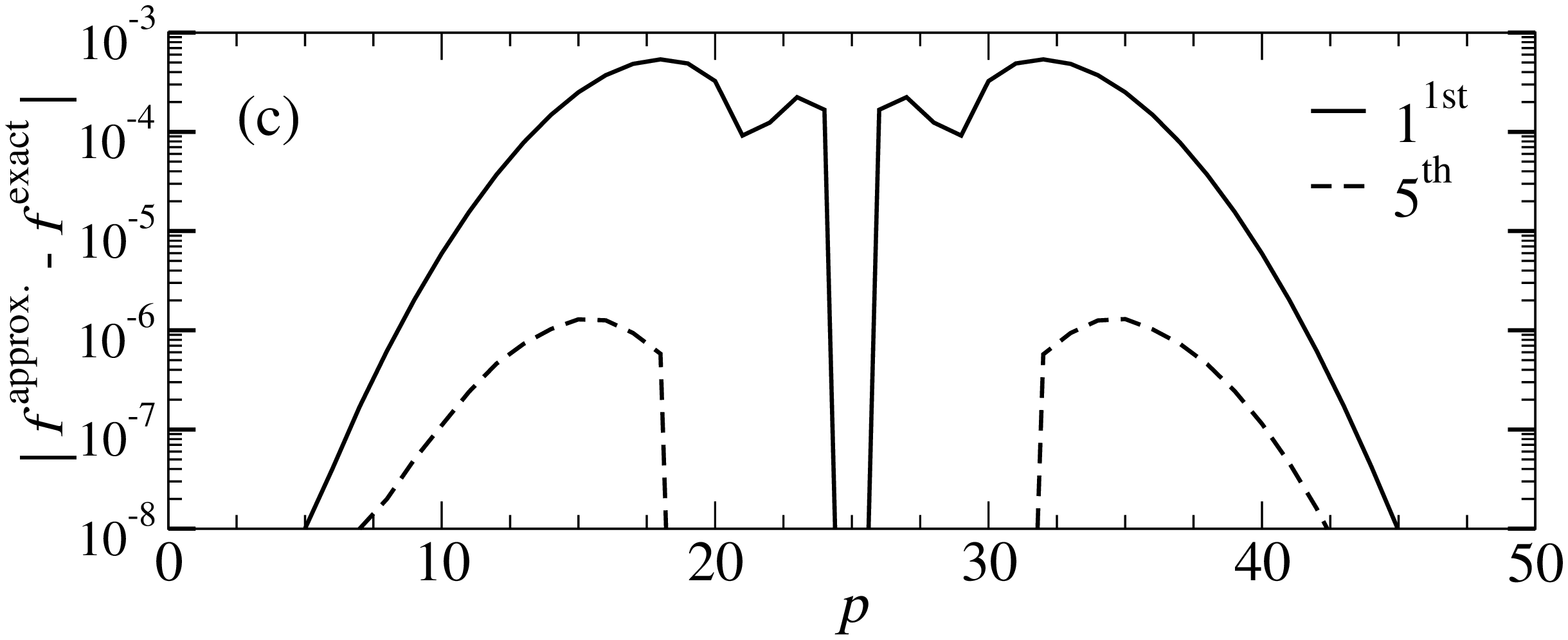}
   \end{center}
  \caption{(Color online) Results for (a) $X(p,q)$ and (b) $f(p,q)$
  as a function of $p$ for $p+q=50$. In both panels, the thick-full
  line corresponds to the ``exact numerical'' result, while thin-full
  and dashed curves are respectively first and fifth order
  approximations. The difference between these latter results
  and the exact ones for $f(p,q)$, almost indistinguishable in (b),
  are shown in (c). Notice that the second term in Eq.~(\ref{X2b})
  vanishes at $p=q$ [implying $f(p,q)=1/2$ for $p=q$] and the
  approximations become exact at this point.
  }
  \label{fig:app}
\end{figure}

We now look for a systematic expansion of $X(p,q)$ in powers of
$(q-p)$ and expand the corresponding coefficients in (non necessarily
integer) powers of $1/r$. In order to perform this expansion, we
rewrite $f(p,q)$ in the form
\begin{equation}
f(p,q)=\frac{1}{2}+\frac{1}{2}\frac{\int_{0}^{1}\left(1-{t^2}\right)^{\frac{r-1}{2}}
\sinh\left[\frac{q-p}{2}\ln\left(\frac{1+t}{1-t}\right)\right]\,dt}
{\int_{0}^{1}\left(1-{t^2}\right)^{\frac{r-1}{2}}
\cosh\left[\frac{q-p}{2}\ln\left(\frac{1+t}{1-t}\right)\right]\,dt}.\label{X2b}
\end{equation}
This expression is formally expanded in powers of $(q-p)$ and the
corresponding coefficients are evaluated for large $r$. This expansion
is then matched with the one obtained from a similar formal expansion
of $X(p,q)$ in Eq.~(\ref{defeq}). This calculation can be carried out
with the help of {\sc Mathematica}, and we finally obtain the fifth
order expansion in $(q-p)$ of $X(p,q)$, which generalizes the first
order result of Eq.~(\ref{order1}). This expansion can be nicely
expressed as an expansion in odd powers of $X_1(p,q)$, with coefficient
having an expansion in integer powers of $1/r$:
\begin{eqnarray}
X(p,q)=&&a_1(r)X_1(p,q)+a_3(r)X_1^3(p,q)+\nonumber\\
&&a_5(r)X_1^5(p,q)+\cdots,\label{X3}
\end{eqnarray}
where $X_1(p,q)$ is given by Eq.~(\ref{X1}), and with the coefficients
\begin{eqnarray*}
a_1(r)&=& 1 - {\frac{1}{12\,r}} - {\frac{19}{160\,{r^2}}} +
   {\frac{155}{2688\,{r^3}}}+\cdots,\\
a_3(r)&=& {\frac{1}{12\,r}} - {\frac{7}{360\,{r^2}}} -
   {\frac{48929}{362880\,{r^3}}} +\cdots,\\
a_5(r)&=& {\frac{43}{1440\,{r^2}}} - {\frac{3253}{362880\,{r^3}}}+\cdots,
\end{eqnarray*}
which were obtained up to third order in $1/r$. As suggested by the
above result, one can indeed show that the expansion of $a_{2l+1}(r)$
in powers of $1/r$ starts at order $l$. Hence, we find that the
expansion of Eq.~(\ref{X3}) is valid for $|q-p|\ll r$ instead of the
naive estimate $|q-p|\ll \sqrt{r}$ which could have been guessed from
the quick first order calculation presented above
Eq.~(\ref{order1}). Since one has $X(p,q)\sim \sqrt{r}$ for $|q-p|\sim
r$, $f(p,q)$ is thus extremely close to 0 ($p>q$) or 1 ($p<q$) in this
regime, with an error exponentially small in $r$. Hence, for all
practical numerical purpose, it is certainly not a serious problem to
have an expansion of $X(p,q)$ limited to $|q-p|\ll r$.

We now briefly illustrate the precision of the above approximate forms
for $f(p,q)$ using the simplest approximation for $X(p,q)$, given in
Eq.~(\ref{X1}), or the fifth order calculation of Eq.~(\ref{X3}).
For the simplest first-order expression of Eq.~(\ref{order1}), the
maximal error is less than $10^{-3}$ for $r>30$, and less than
$10^{-4}$ for $r>275$. For the fifth order approximation, we find a
maximal absolute error which is less than $10^{-5}$, for $r>25$, and
less than $10^{-7}$ for $r>120$. 

In Fig.~\ref{fig:app}(a), we plot $X(p,q)$ as a function of $p$ for
$p+q=50$, for the first and fifth order approximations, and for the
``exact numerical'' $X(p,q)$ obtained after evaluating numerically the
defining integral of $f(p,q)$, and inverting the relation of
Eq.~(\ref{defeq}). The plot of the three corresponding $f(p,q)$ is
presented in Fig.~\ref{fig:app}(b).

The maximal error [$\sim 10^{-3}$, see Fig.~\ref{fig:app}(c)] due to the analytical approximations presented in this Appendix  is well below our Monte Carlo statistical error and for all practical purposes the first-order expression is sufficient. Indeed, only in the case of the results presented in Fig.~\ref{fig:partition}, we faced the case of $r=n\lesssim 30$. Such small values of the SSE expansion order, for which the analytical approximations may become not accurate enough, are only encountered in the case of very small lattices at high temperature. In these cases, a simple pre-computation with a numerical integration of Eq.~(\ref{eq:integral}) for all values of $(m,n)$ can be performed prior to simulations.


\begin{thebibliography}{41}
\bibitem[{con()}]{convention}
\bibinfo{note}{Simulations in the widely adopted convention for the TIM with
  the field in the $x$ direction are particularly efficient using the specific
  algorithm presented in A.W. Sandvik, Phys. Rev. E {\bf 68}, 056701 (2003).
  However, computation of the parity quantum number $P= \prod_{i = 1}^{N}
  \sigma^{x}_{i}\ket{\alpha (\tau = 0)}$ is then highly non-trivial, as the
  $\sigma^x$ are now off-diagonal operators.}
\bibitem[{upd()}]{updates}
\bibinfo{note}{The processes labelled as ``switch-and-reverse" and
  ``switch-and-continue" in Ref.~\onlinecite{syljuasen:02} are always allowed
  in the convention for the TIM adopted here, Eq.~(\ref{eq:tfim01}). See
  Ref.~\onlinecite{syljuasen:02} for details.}
\end{thebibliography}
\end{document}